\documentclass[none]{acmart}
\AtBeginDocument{%
  \providecommand\BibTeX{{%
    \normalfont B\kern-0.5em{\scshape i\kern-0.25em b}\kern-0.8em\TeX}}}

\setcopyright{none}
\copyrightyear{2022}
\acmYear{2022}
\acmDOI{10.1145/3575798}

\acmJournal{TECS}
\acmVolume{1}
\acmNumber{1}
\acmArticle{1}
\acmMonth{12}

\usepackage{scalerel}
\usepackage{tikz}
\usetikzlibrary{svg.path}
\usepackage{xcolor}
\usepackage{amsmath}
\usepackage{amsfonts}
\usepackage{mathtools}
\usepackage[ruled,vlined,noend,linesnumbered]{algorithm2e}
\usepackage{caption}
\usepackage{subcaption}
\usepackage{multirow}
\usepackage{booktabs}
\usepackage{blkarray}
\usepackage[utf8]{inputenc}
\usepackage{fancyhdr}
\usepackage{enumitem}

\newcommand{\latencyReduction}{59.1\% }
\newcommand{\areaIncrease}{17.1\% }
\newcommand{\energyReduction}{60.8\% }
\newcommand{\accuracyIncrease}{1.4\% }

\newcommand{\latencyReductionImageNet}{43.8\% }

\newcommand{\energyReductionImageNet}{11.2\% }
\newcommand{\accuracyIncreaseImageNet}{3.7\% }

\newcommand{\accuracyIncreaseANBA}{1.5\% }
\newcommand{\fpsIncreaseANBA}{34.7$\times$ }
\newcommand{\edpReductionANBA}{11.0$\times$ }
\newcommand{\areaReductionANBA}{4.0$\times$ }

\newcommand{\latencyReductionCvA}{85.7\% }
\newcommand{\latencyReductionCvB}{84.6\% }
\newcommand{\areaReductionCvA}{14.8\% }
\newcommand{\areaReductionCvB}{14.8\% }
\newcommand{\energyReductionCvA}{79.4\% }

\newcommand{\accuracyIncreaseCvA}{4.3\% }
\newcommand{\accuracyIncreaseCvB}{4.2\% }

\def\@fnsymbol#1{\ensuremath{\ifcase#1\or *\or \dagger\or \ddagger\or
   \mathsection\or \mathparagraph\or \|\or **\or \dagger\dagger
   \or \ddagger\ddagger \else\@ctrerr\fi}}
\newcommand{\ssymbol}[1]{^{\@fnsymbol{#1}}}

\begin{document}

\title{CODEBench: A Neural Architecture and Hardware Accelerator Co-Design Framework}

\author{Shikhar Tuli}
\authornote{Both authors contributed equally to this research.}
\email{stuli@princeton.edu}
\orcid{0000-0002-9230-5877}
\author{Chia-Hao Li}
\authornotemark[1]
\email{chli@princeton.edu}
\affiliation{%
  \institution{Princeton University}
  \streetaddress{Department of Electrical and Computer Engineering}
  \city{Princeton}
  \state{NJ}
  \country{USA}
  \postcode{08544}
}

\author{Ritvik Sharma}
\affiliation{%
  \institution{Stanford University}
  \streetaddress{Department of Electrical Engineering}
  \city{Stanford}
  \state{CA}
  \country{USA}
  \postcode{94305}  
}
\email{rsharma3@stanford.edu}

\author{Niraj K. Jha}
\affiliation{%
  \institution{Princeton University}
  \streetaddress{Department of Electrical and Computer Engineering}
  \city{Princeton}
  \state{NJ}
  \country{USA}
  \postcode{08544}
}
\email{jha@princeton.edu}

\thanks{This work was supported by NSF Grant No. CCF-1811109. DOI of published work: https://doi.org/10.1145/3575798}

\renewcommand{\shortauthors}{Tuli et al.}

\begin{abstract}
Recently, automated co-design of machine learning (ML) models and accelerator architectures has attracted 
significant attention from both the industry and academia.
However, most co-design frameworks either explore a limited search space or employ suboptimal exploration 
techniques for simultaneous design decision investigations of the ML model and the accelerator. Furthermore, 
training the ML model and simulating the accelerator performance is computationally expensive.
To address these limitations, this work proposes a novel neural architecture and hardware accelerator co-design 
framework, called CODEBench. It comprises two new benchmarking sub-frameworks, CNNBench and AccelBench, which 
explore expanded design spaces of convolutional neural networks (CNNs) and CNN accelerators. CNNBench leverages an advanced search technique, BOSHNAS, to efficiently train a neural heteroscedastic surrogate model to converge 
to an optimal CNN architecture by employing second-order gradients. AccelBench performs cycle-accurate simulations 
for diverse accelerator architectures in a vast design space. With the proposed co-design method, called
BOSHCODE, our best CNN-accelerator pair achieves \accuracyIncrease higher accuracy on the 
CIFAR-10 dataset compared to the state-of-the-art pair while enabling
\latencyReduction lower latency and \energyReduction lower energy consumption. On the ImageNet dataset, it achieves \accuracyIncreaseImageNet higher Top1 accuracy at \latencyReductionImageNet lower latency and \energyReductionImageNet lower energy consumption. CODEBench outperforms the state-of-the-art
framework, i.e., Auto-NBA, by achieving \accuracyIncreaseANBA higher accuracy and
\fpsIncreaseANBA higher throughput while enabling \edpReductionANBA lower energy-delay product (EDP) and 
\areaReductionANBA lower chip area on CIFAR-10.
\end{abstract}



\maketitle

\section{Introduction}
In the ML community, CNN accelerator design has gained significant popularity recently
\cite{khan2020survey}. Due to the high computational complexity of CNNs, developing a sophisticated hardware accelerator 
is essential to improve the energy efficiency and throughput of the targeted task. Custom accelerators, 
also sometimes called application-specific integrated circuit (ASIC)-based accelerators, outperform
general-purpose processors like central processing units (CPUs) and graphical processing units (GPUs) 
\cite{dadiannao, spring}. However, designing ASIC-based accelerators often comes with its own challenges, such 
as design complexity and long development time. Researchers have constantly strived to design more 
efficient accelerators in terms of throughput, energy consumption, chip area, and model performance. This has led 
to a large number of CNN accelerators, each incorporating different design choices and unique hardware modules to 
optimize certain operations. This requires a plethora of operation-specific hardware modules, sophisticated 
acceleration methods, and many other hyperparameters that define a vast design space in the hardware domain. 
Similarly, on the software side, many design choices for a CNN model result in an enormous design space as well. A challenge that remains is to \emph{efficiently} find an optimal pair of 
software and hardware hyperparameters, in other words, a CNN-accelerator pair that performs well while also 
meeting the required constraints.

The accelerator design space is immense. There are many hyperparameters and one needs to choose their values while 
designing an accelerator for the given application~\cite{VivienneSzeBook}. They include the number and size of processing 
elements (PEs) and on-chip buffers, dataflow, main memory size and type, and many more domain-specific modules, including 
those for sparsity-aware computation and reduced-precision design (more details in Section~\ref{sec:ads}). Similarly, 
the CNN design space is also huge. Many hyperparameters come into play while designing a CNN. They
include the number of layers, convolution type and size, normalization type, pooling type and size, structure of 
the final multi-layer perceptron (MLP) head, activation function, training recipe, and many more (further details 
in Section~\ref{sec:cds}). For a given task, one may search for a particular CNN architecture with the
best performance. However, this architecture may not be able to meet user constraints on power, energy, latency, and chip area.

In the above context, given a hardware module, many works are aimed at tuning the CNN design instead, to optimize 
performance \cite{chamnet}. However, this limits us to the exploration of only the CNN 
design space with no tuning possible in the hardware space. Having a fixed CNN, one could also optimize the hardware (called \emph{automatic accelerator synthesis}) in the context where searching the CNN space may be too expensive, especially for large datasets. This hardware optimization, however, requires a thorough knowledge of the architecture and 
design of ASIC-based accelerators, leading to long design cycles. On the other hand, exploring 
the CNN design space falls under the domain of neural architecture search (NAS). Advancements 
in this domain have led to many NAS algorithms, including those employing reinforcement learning (RL), Bayesian optimization, structure adaptation, etc.~\cite{reinforce, gp_bo, nest}. However, these approaches have many limitations including suboptimal convergence, slow regression performance, and are limited to fixed training recipes (details in 
Section~\ref{sec:cnnbench}). Tuli et al.~\cite{flexibert} recently proposed a model for NAS in the space of transformer 
ML models~\cite{vaswani}. It overcomes many of these challenges by leveraging a heteroscedastic surrogate model to search 
for the model's design decisions and its training recipe. However, this technique is not amenable to co-design between 
the two (namely, the accelerator and CNN) design spaces (details in Section~\ref{sec:boshcode}).

This work also shows that a one-sided search leads to suboptimal CNN-accelerator pairs. This has led to recent advancements in 
\emph{co-design} of both the software and hardware. Researchers often leverage RL 
techniques to search for an optimal CNN-accelerator pair \cite{hw_sw_co-exp, bobw, nahas}. However, most co-design
works only use local search (mutation and crossover) and/or have limited search spaces, e.g., they only 
search over selected field-programmable gate arrays (FPGAs) or microcontrollers \cite{bobw, mcunet}.
Some works have also leveraged differentiable search of the CNN architecture~\cite{edd_dac,
dance_dac}. However, recent surveys have shown that these methods are much slower than surrogate-based methods and fail
to explore potential clusters with higher performance models~\cite{nasbench301}. Limited search spaces, as shown by
some very recent works~\cite{flexibert, auto_nba}, often lead to suboptimal neural network models and even
CNN-accelerator pairs (or, in general, the combination of the hardware architecture and the software algorithm). Thus,
expanding existing design spaces in both the hardware and the software regimes is necessary. However, blindly
growing these design spaces further prolongs design times and exponentially increases compute resource requirements.

To tackle the above challenges, we propose a framework for comprehensively and simultaneously exploring 
\emph{massive} CNN architecture 
and accelerator design domains, leveraging a novel co-design workflow to obtain an optimal CNN-accelerator pair 
for the targeted application that meets user-specified constraints. These could include not just edge applications with
highly constrained power envelopes, but also server applications where model accuracy is of utmost importance. 

Our optimal CNN-accelerator pair outperforms the state-of-the-art pair~\cite{spring}, achieving \accuracyIncrease higher model accuracy on the CIFAR-10 dataset~\cite{cifar10, imagenet}
while enabling \latencyReduction lower latency and \energyReduction lower energy consumption, with only \areaIncrease increase in
chip area. This pair also achieves \accuracyIncreaseImageNet higher Top1 accuracy on the ImageNet dataset while incurring \latencyReductionImageNet lower latency and \energyReductionImageNet lower energy (with the same increase in chip area). Experiments with our expanded design spaces that include popular CNNs and accelerators, using our proposed framework, show an improvement of 
\accuracyIncreaseANBA in model accuracy on the CIFAR-10 dataset, while enabling \edpReductionANBA lower EDP and \fpsIncreaseANBA higher throughput, with \areaReductionANBA lower area for our CNN-accelerator pair relative to a state-of-the-art co-design framework, namely Auto-NBA~\cite{auto_nba}. We plan to release trained CNN models, accelerator architecture simulations, and our framework to enable future benchmarking.

\begin{figure*}[t]
    \centering
    \includegraphics[width=\linewidth]{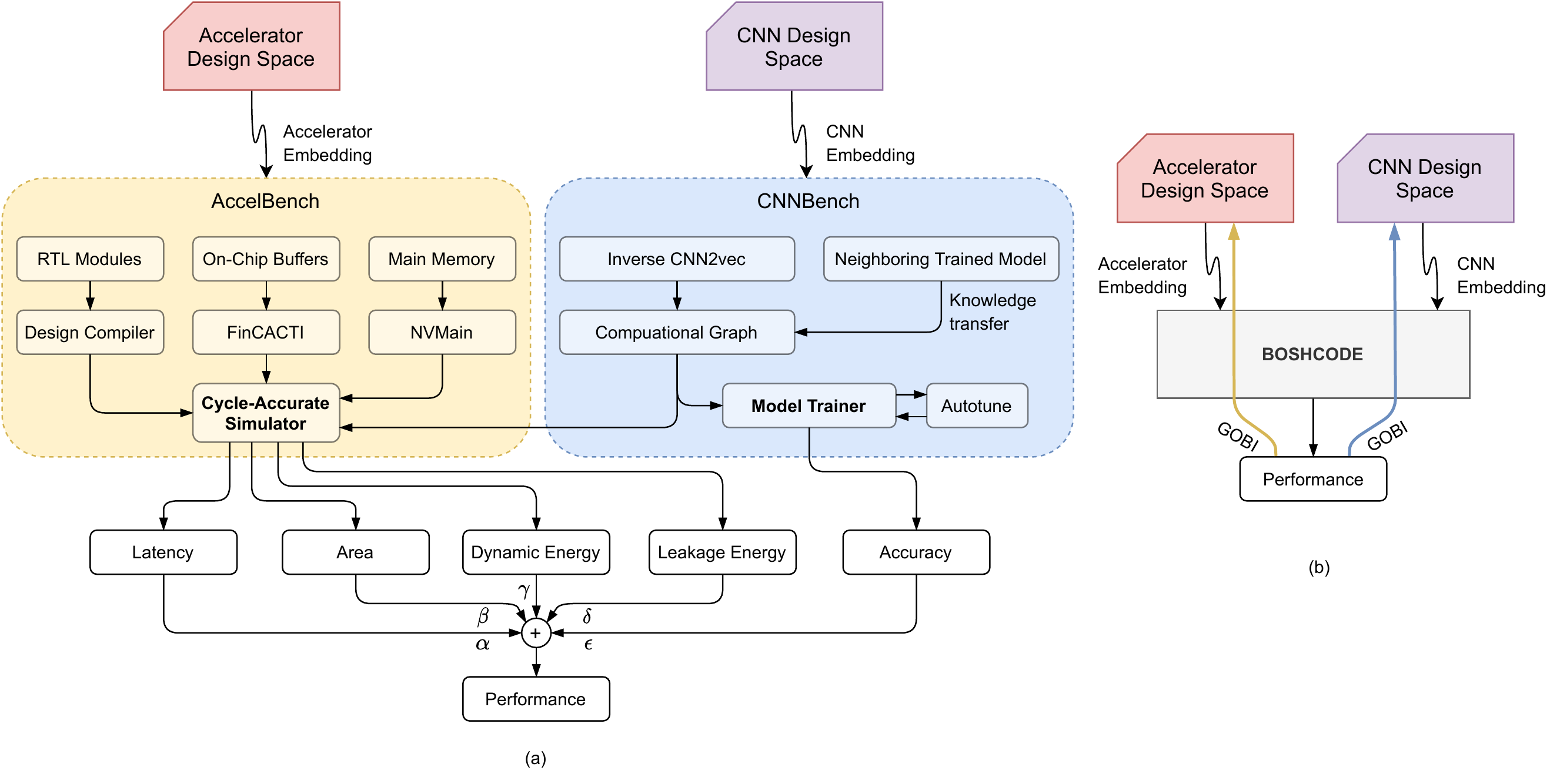}
    \caption{CODEBench pipeline that includes CNNBench, AccelBench, and a novel co-design method, BOSHCODE: (a) The CNN and accelerator design spaces are sampled for novel 
CNN-accelerator pairs that are simulated using the CNNBench and AccelBench frameworks. (b) BOSHCODE learns 
a surrogate function from these design spaces to predict the performance of each CNN-accelerator pair, implementing GOBI to predict the next pair to be simulated in the active learning framework.}
    \label{fig:code_pipeline}
\end{figure*}

The main contributions of this article are summarized next.
\begin{itemize} 
    \item We expand on previously proposed CNN design spaces and present a new
tool, called CNNBench, to characterize the vast design space of CNN architectures that includes a diverse set 
of supported convolution operations, unlike any previous work~\cite{edd_dac, naas, nasbench101}. We propose \texttt{CNN2vec} 
that employs similarity measures to compare computational graphs of CNN models in order to obtain a dense embedding 
that captures architecture similarity reflected in the Euclidean space. We also leverage a new NAS 
technique, \underline{B}ayesian \underline{O}ptimization 
using \underline{S}econd-order Gradients and \underline{H}eteroscedastic Surrogate Model for 
\underline{N}eural \underline{A}rchitecture \underline{S}earch (BOSHNAS), for searching our expanded design space. 
CNNBench also leverages similarity between neighboring CNN computational graphs for weight transfer to 
speed up the training process. Due to the massive design space, along with \texttt{CNN2vec} embeddings, weight transfer from previously trained neighbors, and simultaneous optimization of the CNN architecture and the training recipe, CNNBench achieves state-of-the-art performance while limiting the number of search iterations compared to previous works~\cite{bananas}.
    \item We survey popular accelerators proposed in the literature and encapsulate their design decisions in a 
unified framework. This gives rise to a benchmarking tool, AccelBench, that runs inference of 
CNN models on any accelerator within the design space, employing cycle-accurate simulations. AccelBench incorporates accelerators with diverse memory configurations that one could use for future benchmarking, rather than using traditional ASIC templates~\cite{nasaic, auto_nba}. With the goal to reap the benefits of vast design spaces~\cite{flexibert, auto_nba}, AccelBench is the first benchmarking tool for diverse ASIC-based accelerators, supporting variegated design decisions in 
modern accelerator deployments. Unlike previous works that use FPGAs or off-the-shelf ASIC templates, it builds accelerator designs from the ground up, mapping each CNN in a modular and efficient fashion. It supports $2.28 \times 10^8$ unique accelerators, a design space much more extensive than 
investigated in any previous work.
    \item In order to \emph{efficiently} search the proposed massive design space, we present a novel co-design method, \underline{B}ayesian \underline{O}ptimization 
using \underline{S}econd-order Gradients and \underline{H}eteroscedastic Surrogate Model for 
\underline{Co}-\underline{De}sign of CNNs and Accelerators (BOSHCODE). To make the search of such a vast design space 
possible, BOSHCODE incorporates numerous novelties, including a hierarchical search technique that gradually increases the 
granularity of hyperparameters searched, a neural network surrogate-based model that leverages gradients to the input for 
reliable query prediction, and an active learning pipeline that makes the search more efficient. Here, by gradients to the input, we mean the gradients to the CNN-accelerator pair simulated in the next iteration of the active learning loop. BOSHCODE is a fundamental pillar for the joint exploration of vast hardware-software design spaces. CODEBench, our proposed framework, combines CNNBench (which trains and obtains the accuracy of any queried CNN),
AccelBench (which simulates the hardware performance of any queried accelerator architecture), and the 
proposed co-design method, BOSHCODE, to find the optimal CNN-accelerator pair, given a set of
user-defined constraints. Fig.~\ref{fig:code_pipeline} presents an overview of the proposed framework. Fig.~\ref{fig:code_pipeline}(a) shows how the CNNBench and AccelBench simulation pipelines output the
performance values for every CNN-accelerator pair. Fig.~\ref{fig:code_pipeline}(b) shows how BOSHCODE 
learns a surrogate model for this mapping from all CNN-accelerator pairs to their simulated performance values. CNNBench 
trains the CNN model to obtain model accuracy and feeds the checkpoints to AccelBench that obtains other performance measures using a cycle-accurate simulator.
\end{itemize}

The rest of the article is organized as follows. Section~\ref{sec:back_rel} discusses the CNN and accelerator 
design spaces and highlights the advantages of co-design over one-sided optimization approaches. Section~\ref{sec:code} 
presents the co-design framework that uses BOSHCODE to search for an optimal CNN-accelerator pair. 
Section~\ref{sec:exp} describes the experimental setup and baselines considered. Section~\ref{sec:res} discusses 
the results. Finally, Section~\ref{sec:conc} concludes the article\footnote{The associated code for CNNBench is available at \url{https://github.com/jha-lab/cnn_design-space}. For the entire CODEBench pipeline, the code is available at \url{https://github.com/jha-lab/codebench}.}.

\section{Background and Related Work}
\label{sec:back_rel}

In this section, we present background material and related works in the fields of CNN synthesis 
and accelerator design. 

\subsection{CNN Design Space}
\label{sec:cds}

First, we discuss design choices in the CNN design space. We show how various CNN building blocks contribute to 
its performance or efficiency. We then discuss popular NAS techniques and hardware-aware NAS.

\subsubsection{Popular CNN Architectures}

The CNN design space we consider encompasses popular CNNs of various sizes. The first is LeNet~\cite{lenet}, one of the earliest successful CNNs. AlexNet was the first CNN to propose grouped 
convolutions~\cite{alexnet}. MobileNet introduced bottleneck connections that enable parameter 
reduction~\cite{mobilenetv2} (we use MobileNet-V2 as our baseline for subsequent discussions). ResNet proposed residual connections that aided backpropagation of gradients as 
networks grew deeper~\cite{resnet}. ShuffleNet presented the shuffled convolution, an extension to grouped 
convolution, that introduced the \emph{channel shuffle} operation for improved regularization and thus 
efficiency~\cite{shufflenet}. Other target CNNs we can represent within our design space include 
GoogleNet, Xception, SqueezeNet, DenseNet, 
and VGG and EfficientNet families~\cite{khan2020survey, googlenet, xception, squeezenet, densenet, vgg, efficientnet}.

\subsubsection{Neural Architecture Search}

NAS incorporates various search techniques that algorithmically find new neural architectures within a 
pre-defined space based on a given objective~\cite{nas_survey}. Prior works have implemented NAS using a variety 
of techniques. A popular approach is to use an RL algorithm, REINFORCE, that is superior to other 
tabular approaches~\cite{reinforce}. Other techniques include Gaussian-process based Bayesian optimization 
(GP-BO)~\cite{gp_bo}, structure adaptation methods based on grow-and-prune synthesis~\cite{nest}, differentiable and proxy-based architecture search~\cite{darts, proxyless_nas}, local search techniques 
including mutation, and evolutionary search (ES)~\cite{white2020local}. 
Recently, NAS has also seen the application of surrogate models for CNN performance prediction owing
to various drawbacks of the methods mentioned above, including high compute cost, slow convergence, etc.~\cite{nasbench301}. 
Exploiting surrogate performance results in training much fewer models to predict the accuracy of CNNs in the entire 
design space under some confidence constraints. However, these predictors are computationally expensive to train, bottlenecking the search process. NASBench-301~\cite{nasbench301} uses a graph isomorphism network (GIN) that regresses performance on graphs. Recent works show that it is slow enough to bottleneck the search process~\cite{flexibert}.

One of the state-of-the-art NAS techniques for CNNs, BANANAS~\cite{bananas}, implements Bayesian Optimization (BO) using a neural network (NN) 
surrogate and predicts performance uncertainty using ensemble networks that are compute-heavy. BANANAS 
uses mutation and crossover to get the set of current best-performing models and obtains the next best-predicted 
model in this local space. Instead, we leverage BOSHNAS~\cite{flexibert} to efficiently search for the following 
query in the \emph{global} space curated by our CNNBench framework. Due to random cold restarts, BOSHNAS can search over diverse models in the 
architecture space. Moreover, BANANAS also uses path embeddings that perform suboptimally in 
search over a diverse space of CNN models \citep{nasgem}. To address this, we propose a novel embedding, 
\texttt{CNN2vec}, which is dense and thus more amenable to surrogate modeling.

\subsubsection{Benchmarking for NAS}

Previously, different works on NAS used disparate training pipelines, search spaces, and hyperparameter sets, 
and did not evaluate other methods under comparable settings. To address these issues, recent works proposed various NAS benchmarks~\cite{nasbench101, nasbench301}. For instance, BANANAS uses the NASBench-101 dataset to empirically 
show how it improves upon other NAS algorithms~\cite{bananas}. However, these benchmarks also have their limitations. 
NASBench-101~\cite{nasbench101} is only a tabular dataset of limited size and its results do not transfer well 
to realistic search spaces~\cite{white2020local}. As explained above, NASBench-301~\cite{nasbench301} is a surrogate NAS benchmark 
that uses GIN as its performance predictor. However, a GIN is slow to evaluate and  
chooses a static training recipe for every CNN 
model in the design space. In other words, previous works only consider the \emph{epistemic}
uncertainty and not the \emph{aleatoric} uncertainty in predictions. The former, also called reducible uncertainty,
arises from a lack of knowledge or information, and the latter, also called irreducible uncertainty, refers to the
inherent variation in the system to be modeled (in our context, this corresponds to the change in performance of a given 
CNN architecture with different training recipes).

Another work in this direction trains a large network and derives sub-networks for different tasks and targeted 
hardware platforms~\cite{once_for_all}. However, it only considers pyramidal structures, has a limited 
library of building blocks, uses a static training recipe, and employs sparse embeddings that are not amenable 
to performance predictors. CNNBench, our proposed benchmarking framework, generates a vast 
design space of CNNs using an expanded space of building blocks and decouples the surrogate model from the graph 
modeling process. It does so using the proposed \texttt{CNN2vec} embeddings to speed up the search process (details in 
Section~\ref{sec:cnnbench}).

\subsubsection{Hardware-aware NAS}
\label{sec:hw_nas}

NAS alone is hardly useful if one cannot run the best-performing CNNs on the hardware at hand. Recent works 
have thus focused on \emph{hardware-aware} NAS, which constrains and directs CNN search based on the targeted 
hardware platform~\cite{hw_nas_survey}. ChamNet proposed accuracy and resource (latency and energy) predictors 
and leveraged GP-BO to find the optimal CNN architecture for a given platform~\cite{chamnet}. However, it used 
ES to search for a variant of the base NN that may miss other clusters of high-performing networks, which is
a drawback of \emph{elitist} algorithms~\cite{nasbench301,bananas,elitist}. Furthermore, it trains a separate latency predictor for every hardware 
platform, which is a time-consuming endeavor. ProxylessNAS used a proxy-based differentiable search technique. However, it only 
works with latency and not other hardware performance measures, and is also slower than state-of-the-art surrogate-based methods. 
Many other works have strived to address hardware-aware NAS, but 
under a limited scope, often restricted to commercial edge devices, FPGAs, and off-the-shelf ASIC 
templates~\cite{chamnet, hw_sw_co-exp, hw_nas_bench}.

\subsection{Accelerator Design Space}
\label{sec:ads}

Next, we present design choices in the accelerator design space. Many ASIC-based accelerators were 
proposed in previous works to tackle the task of CNN acceleration, i.e., attain the highest model accuracy while also achieving high 
energy and area efficiency. These accelerators address the task with 
various methodologies and custom hardware modules. We present a brief survey of some seminal works on accelerator 
design and show how they motivate the need for a vast design space and automated search for an optimal architecture.

\subsubsection{Popular Accelerators}

The Eyeriss v1 and Eyeriss v2~\cite{eyerissv2} accelerators reduce memory accesses through an efficient row stationary (RS) data reuse strategy (also called a \emph{dataflow}). Other dataflows have been investigated, including input, output, and weight stationary (IS, OS, and WS, respectively) dataflows~\cite{VivienneSzeBook}. Eyeriss v2 further exploits sparsity by directly processing input feature maps and filter weights in the compressed sparse column (CSC) format. DianNao~\cite{diannao}, DaDianNao~\cite{dadiannao}, and ShiDianNao~\cite{shidiannao} form another family of accelerators that first enabled sophisticated buffer designs, following the OS dataflow~\cite{VivienneSzeBook}. Cambricon-X~\cite{cambricon-x} and Cambricon-S~\cite{cambricon-s} are successors of the DianNao series, encapsulating dedicated indexing modules to exploit sparsity in NNs. Cnvlutin~\cite{cnvlutin} is another accelerator that focuses on eliminating the ineffectual data in the input feature maps in CNNs and uses the zero-free neuron array format (ZNAf) to compress data. Finally, SPRING~\cite{spring}, a state-of-the-art accelerator, leverages 3D non-volatile memory for increased bandwidth, exploits reduced-precision techniques for data movement efficiency, and OS dataflow for data reuse.

\setlength{\tabcolsep}{3pt}
\begin{table*}[!t]
    \centering
    \caption{The hyperparameters and design choices of previous works on CNN accelerators.}
    \resizebox{\textwidth}{!}{
    \begin{tabular}{@{}lllllllllllllll@{}}
    \toprule
        Accelerator & Technology & Clock & Area & Number of & Number of & Number of & Weight & Activation & Other & Precision & Main & Main & Dataflow & Sparsity- \\
        Name & Node & Rate & ($mm^2$) & PEs & MAC Units & Multipliers & Buffer & Buffer & Buffer & & Memory & Memory &  & Aware\\ 
        & & (MHz) & & & Per PE & Per MAC & Size & Size & Size & & Interface & Type & & Scheme\\
         & & & & & & Unit & & & & & & & & \\
        \midrule
        \multirow{2}{*}{Eyeriss} & 65nm &  \multirow{2}{*}{200} & \multirow{2}{*}{16} & \multirow{2}{*}{168} & \multirow{2}{*}{1} & \multirow{2}{*}{1} & \multicolumn{2}{c}{\multirow{2}{*}{108 KB}} & \multirow{2}{*}{87 KB} & \multirow{2}{*}{int16} & \multirow{2}{*}{2D} & \multirow{2}{*}{DRAM} & \multirow{2}{*}{RS} & Data \\
         & CMOS & & & & & & & \multicolumn{2}{c}{} & & & & & Gating\\
        \multirow{2}{*}{Eyeriss V2} & 65nm & \multirow{2}{*}{200} & \multirow{2}{*}{37} & \multirow{2}{*}{192} & \multirow{2}{*}{1} & \multirow{2}{*}{2} & \multicolumn{2}{c}{\multirow{2}{*}{192 KB}} & \multirow{2}{*}{105 KB} & \multirow{2}{*}{int8} & \multirow{2}{*}{2D} & \multirow{2}{*}{DRAM} & \multirow{2}{*}{RS} & CSC \\
         & CMOS & & & & & & & \multicolumn{2}{c}{} & & & & & Compression\\
        \multirow{2}{*}{DianNao} & 65nm & \multirow{2}{*}{980} & \multirow{2}{*}{3.02} & \multirow{2}{*}{1} & \multirow{2}{*}{16} & \multirow{2}{*}{16} & \multirow{2}{*}{32 KB} & \multirow{2}{*}{4 KB} & \multirow{2}{*}{8 KB} & \multirow{2}{*}{int16} & \multirow{2}{*}{2D} & \multirow{2}{*}{DRAM} & \multirow{2}{*}{OS} & \multirow{2}{*}{N/A} \\
         & CMOS & & & & & & & & & & & & & \\
        DaDianNao & 28nm & \multirow{2}{*}{606} & \multirow{2}{*}{67.73} & \multirow{2}{*}{16} & \multirow{2}{*}{16} & \multirow{2}{*}{16} & \multirow{2}{*}{32 MB} & \multirow{2}{*}{4 MB} & \multirow{2}{*}{N/A} & \multirow{2}{*}{int16} & \multirow{2}{*}{2D} & \multirow{2}{*}{DRAM} & \multirow{2}{*}{OS} & \multirow{2}{*}{N/A} \\ 
        (1 node) & CMOS & & & & & & & & & & & & & \\
        \multirow{2}{*}{ShiDianNao} & 65nm & \multirow{2}{*}{1000} & \multirow{2}{*}{4.86} & \multirow{2}{*}{64} & \multirow{2}{*}{1} & \multirow{2}{*}{1} & \multirow{2}{*}{128 KB} & \multirow{2}{*}{128 KB} & \multirow{2}{*}{32 KB} & \multirow{2}{*}{int16} & \multirow{2}{*}{2D} & \multirow{2}{*}{DRAM} & \multirow{2}{*}{OS} & \multirow{2}{*}{N/A} \\ 
         & CMOS & & & & & & & & & & & & & \\
        \multirow{2}{*}{Cambricon-X} & 65nm & \multirow{2}{*}{1000} & \multirow{2}{*}{6.38} & \multirow{2}{*}{16} & \multirow{2}{*}{1} & \multirow{2}{*}{16} & \multirow{2}{*}{32 KB} & \multirow{2}{*}{16 KB} & \multirow{2}{*}{4 KB} & \multirow{2}{*}{int16} & \multirow{2}{*}{2D} & \multirow{2}{*}{DRAM} & \multirow{2}{*}{WS} & Step \\ 
         & CMOS & & & & & & & & & & & & & Indexing\\
        \multirow{2}{*}{Cambricon-S} & 65nm & \multirow{2}{*}{1000} & \multirow{2}{*}{6.73} & \multirow{2}{*}{16} & \multirow{2}{*}{1} & \multirow{2}{*}{16} & \multirow{2}{*}{32 KB} & \multirow{2}{*}{16 KB} & \multirow{2}{*}{1 KB} & \multirow{2}{*}{int16} & \multirow{2}{*}{2D} & \multirow{2}{*}{DRAM} & \multirow{2}{*}{WS} & Direct \\ 
         & CMOS & & & & & & & & & & & & & Indexing\\
        \multirow{2}{*}{Cnvlutin} & 65nm & \multirow{2}{*}{1000} & \multirow{2}{*}{71} & \multirow{2}{*}{16} & \multirow{2}{*}{16} & \multirow{2}{*}{16} & \multirow{2}{*}{32 MB} & \multirow{2}{*}{N/A} & \multirow{2}{*}{4 MB} & \multirow{2}{*}{int16} & \multirow{2}{*}{2D} & \multirow{2}{*}{DRAM} & \multirow{2}{*}{OS} & ZFNAf \\ 
         & CMOS & & & & & & & & & & & & & Compression\\
        \multirow{2}{*}{SPRING} & 14nm & \multirow{2}{*}{700} & \multirow{2}{*}{151} & \multirow{2}{*}{64} & \multirow{2}{*}{72} & \multirow{2}{*}{16} & \multirow{2}{*}{24 MB} & \multirow{2}{*}{12 MB} & \multirow{2}{*}{4 MB} & \multirow{2}{*}{int20} & Monoli- & \multirow{2}{*}{RRAM} & \multirow{2}{*}{OS} & Binary- \\
         & FinFET & & & & & & & & & & thic 3D & & & Mask\\
        \bottomrule
    \end{tabular}}
    \label{tbl:accelerators}
\end{table*}

Table~\ref{tbl:accelerators} lists the hyperparameters and design choices of the aforementioned
accelerators. These works propose various methodologies, however, each with its overhead. 
Thus, performing a rigorous comparison of these methodologies is important~\cite{VivienneSzeBook}. This can
be facilitated by benchmarking various accelerator designs to trade off improvements offered by them with the
overheads incurred. Moreover, one needs to ascertain the values of many hardware hyperparameters to 
design an efficient accelerator, such as the number of PEs, size of on-chip buffers, etc. These hyperparameters allow ASIC-based accelerators to be highly 
customized for targeted applications and achieve high efficiency. However, the development of such accelerators often
comes at the cost of long design cycles and requires considerable computational resources and domain expertise. Hence, many works have strived to automate the accelerator design process. 

\subsubsection{Automatic Accelerator Synthesis}

The accelerator architecture must adequately match the targeted CNN 
model to achieve optimal CNN acceleration and enable maximum utility and parallelism. Several works target automatic accelerator synthesis methods to explore their self-defined design spaces systematically \cite{aladdin, softwaredefined}. 
However, they miss the benefits and improvements obtained by exploring the CNN design space simultaneously. This motivates the need for a co-design framework that explores the CNN and accelerator architecture design spaces 
simultaneously.

\subsection{Hardware-software Co-design}

Due to the benefits of both NAS and automated accelerator design, hardware-software co-design methods 
explore an optimal CNN-accelerator pair. Many RL-based methods target co-design~\cite{hw_sw_co-exp, bobw}. However, these frameworks are only applicable to FPGA-based accelerators. Zhang et al.~\cite{fpga_design} propose a rigorous design space of FPGA-based accelerators. However, it does not consider various memory types (e.g., off-chip DRAM, HBM, and RRAM), which limits its applicability. Recent works factor this in with a memory bandwidth parameter~\cite{fpga_design, accelerator_bo}. However, the memory configuration also affects the memory controller and area (for monolithic adoption), which previous works do not target. Finally, these works do not consider the batch size (and the corresponding tiling parameters) that affect both model accuracy and hardware performance.
NAHAS~\cite{nahas}, another RL-based exploration method, performs co-design based on the design space of ASIC-based 
accelerators. Nevertheless, the framework only focuses on a small CNN design space with limited CNN 
architectural hyperparameters. Moreover, it does not take the energy consumption of the accelerators and the cost of 
the main memory system into account when performing the evaluation. EDD~\cite{edd_dac} and DANCE~\cite{dance_dac} instead exploit differentiable search of the CNN model. Evolution-based and differentiable search algorithms only perform a local search around the 
parent designs, thus missing out on other potentially superior CNN-accelerator pairs in the design space~\cite{nasbench301}. Gibbon~\cite{gibbon} leverages ES (drawbacks discussed in Section~\ref{sec:hw_nas}) on a design space of in-memory accelerators, although with an adaptive parameter priority pipeline. However, it is limited to in-memory accelerators, 
does not employ uncertainty in exploration, and does not tune the training recipe for the chosen CNN model. Another recent 
work, NAAS \cite{naas}, leverages an evolution-based algorithm to explore the ASIC-based hardware space together with 
compiler mapping strategies. 
The above shortcomings motivate the development of a thorough and sophisticated co-design method 
based on comprehensive CNN and accelerator design spaces. Next, we present our proposed solution that tackles 
these challenges.

\section{The CODEBench Framework}
\label{sec:code}

\begin{figure}[t]
    \centering
    \includegraphics[width=0.5\linewidth]{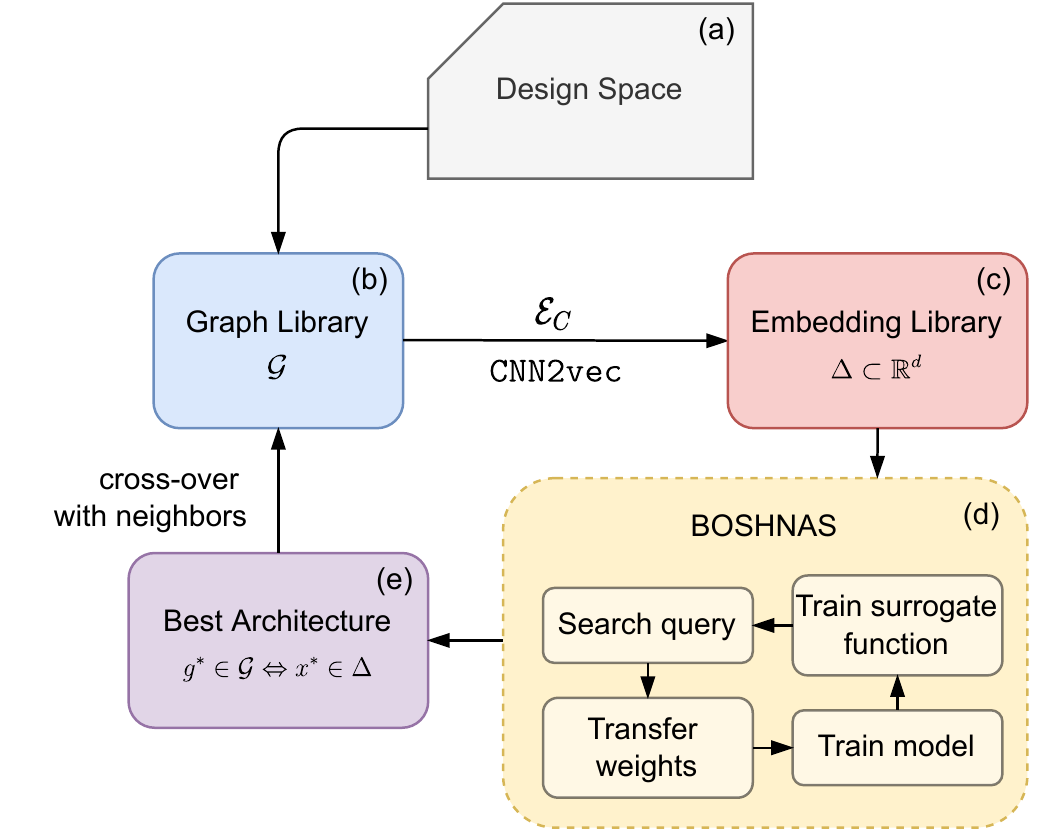}
    \caption{Flowchart of different parts of the CNNBench pipeline.}
    \label{fig:boshnas_flowchart}
\end{figure}

We now discuss the proposed co-design pipeline in detail. 

\subsection{CNNBench}
\label{sec:cnnbench}

We discuss different parts of the CNNBench pipeline (summarized in Fig.~\ref{fig:boshnas_flowchart}) next.

\subsubsection{Design Space and Model Cards}

The CNNBench design space (represented by Fig.~\ref{fig:boshnas_flowchart}(a)) is formed by \emph{computational graphs} for a diverse set of CNN architectures. All models in CNNBench are 
combinations of one or more operations. These operations include not only traditional convolutions, but also 
many more, in light of recent state-of-the-art CNN architectures. CNNBench supports \textit{2D convolution} 
operations of various kernel sizes ($1 \times 1$ to $11 \times 11$) with different number of channels (4 to 
8256), number of groups (4 or 8), padding (1 to 3), along with \textit{ReLU}, \textit{SiLU}
\cite{silu}, and other activation functions (further details in Section~\ref{sec:exp_cds}). We also support \textit{depth-wise separable convolutions} 
\cite{mobilenetv2}, \textit{3D convolutions}, and \textit{transposed convolutions}~\cite{khan2020survey}. 
Each convolutional block in our design space is a combination of the convolution operation itself, 
followed by a \textit{batch-normalization} step and the activation function \cite{nasbench101}. 

Unlike the NASBench-101 dataset~\cite{nasbench101}, our design space also includes \textit{dropout} operations \cite{dropout}, \textit{channel-shuffle} operation with different group sizes \cite{shufflenet} along with 
max and average \textit{pooling} with different kernel sizes, padding, and strides. In line with the 
EfficientNet family, we also include bilinear \textit{upsampling} operations to support 
very deep networks that would otherwise suffer from vanishing spatial dimensions~\cite{efficientnet}. 
This expanded set of operations makes the design space flexible enough for more accelerator-friendly models with respect to efficiency and hardware capacity.

\subsubsection{Block-level Computational Graphs}

\begin{figure}[t]
    \centering
    \includegraphics[width=0.7\linewidth]{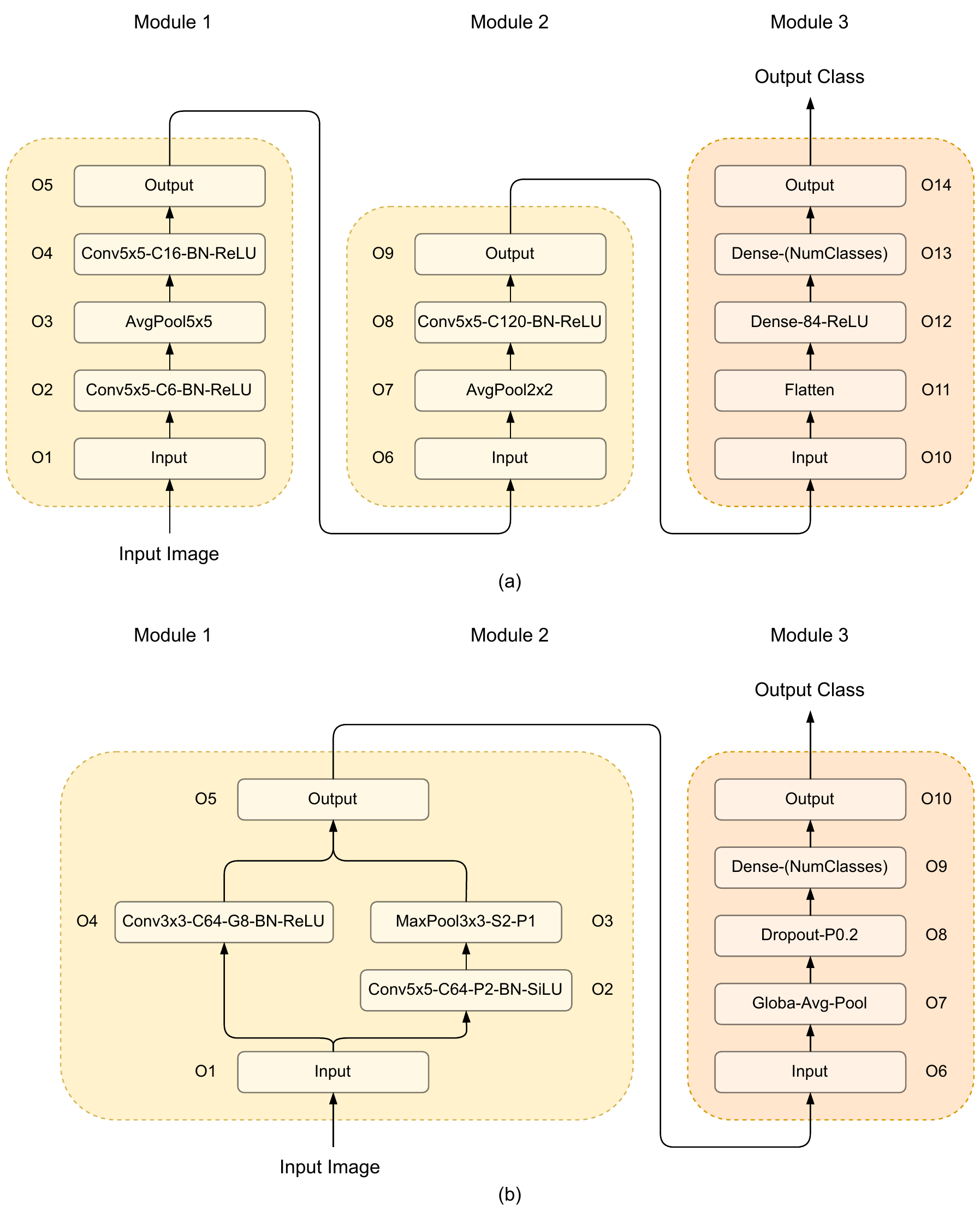}
    \caption{Computational graph of (a) LeNet and (b) a complex CNN in the CNNBench design space. \emph{Dense-(NumClasses)} is a feed-forward 
layer based on the number of output classes for the given dataset~\cite{cifar10, imagenet}. Stride and padding default to 1 if not 
mentioned.}
    \label{fig:comp_graph_cnnbench}
\end{figure}

Using the operation blocks defined above, we create various possible CNN architectures in the form of 
computational graphs. These forward flows of operations in the network determine the computational graph. 
Fig.~\ref{fig:comp_graph_cnnbench}(a) shows the computational graph of LeNet~\cite{lenet}. Using the set of 
permissible operation blocks, we create all possible computational graphs for the design space (represented by Fig.~\ref{fig:boshnas_flowchart}(b); details in 
Section~\ref{sec:exp}). Fig.~\ref{fig:comp_graph_cnnbench}(b) shows the computational graph of another possible CNN in our design space. We create each computational graph in the form of \emph{modules}, where each module is 
a set of interconnected blocks with one input and one output. For instance, in Fig.~\ref{fig:comp_graph_cnnbench}(b), 
there are two modules, one for the convolutional operations (yellow) and the other for the MLP head (orange). We stack multiple such modules to create deep and complex CNN architectures in our design space.

These modules are serially connected to form the computational graphs in 
the design space. This restricts the creation of highly complex graphs. For our design space, we limit the size 
of convolutional modules to five operations (including \textit{input} and \textit{output} blocks) with a maximum 
of eight edges in the graph per module.
The final head can 
have up to eight operations that are sequentially connected. Then, using the set of computational blocks, we create the design space of CNN 
architectures in this modular format. 

\subsubsection{Levels of Hierarchy}
\label{sec:hierarchy}

Creating graphs using the above approach may lead to an extremely large design space. To make the 
exploration more tractable, we propose a hierarchical search method that searches the design decisions in an increasingly granular fashion. One can consider each CNN architecture created in the 
design space to be composed of multiple stacks of modules. In this context, a stack is just a 
serial connection of modules where each module in the stack is inherently the same. For instance, if the number 
of modules in a stack is 10, i.e., $s=10$, in a CNN with 31 modules, the first 10 modules would be the 
same, then the next 10, and the next 10 after that, after which there would be a module for the MLP head. To go 
from one level of the hierarchy to the next, we consider a design space constituted by a finer-grained 
neighborhood of these models. We derive the neighborhood by pairwise crossover between the best-performing 
models and their neighbors in the current level of the hierarchy where the number of modules per stack is $s$ 
(found using a graph-similarity method discussed later), in a space where the number of modules per stack is $s/2$ (or any integer divisor of $s$). We explain this crossover in detail next.

\subsubsection{Crossover between CNN Models}
\label{sec:crossover}

\begin{figure}[t]
    \centering
    \includegraphics[width=0.5\linewidth]{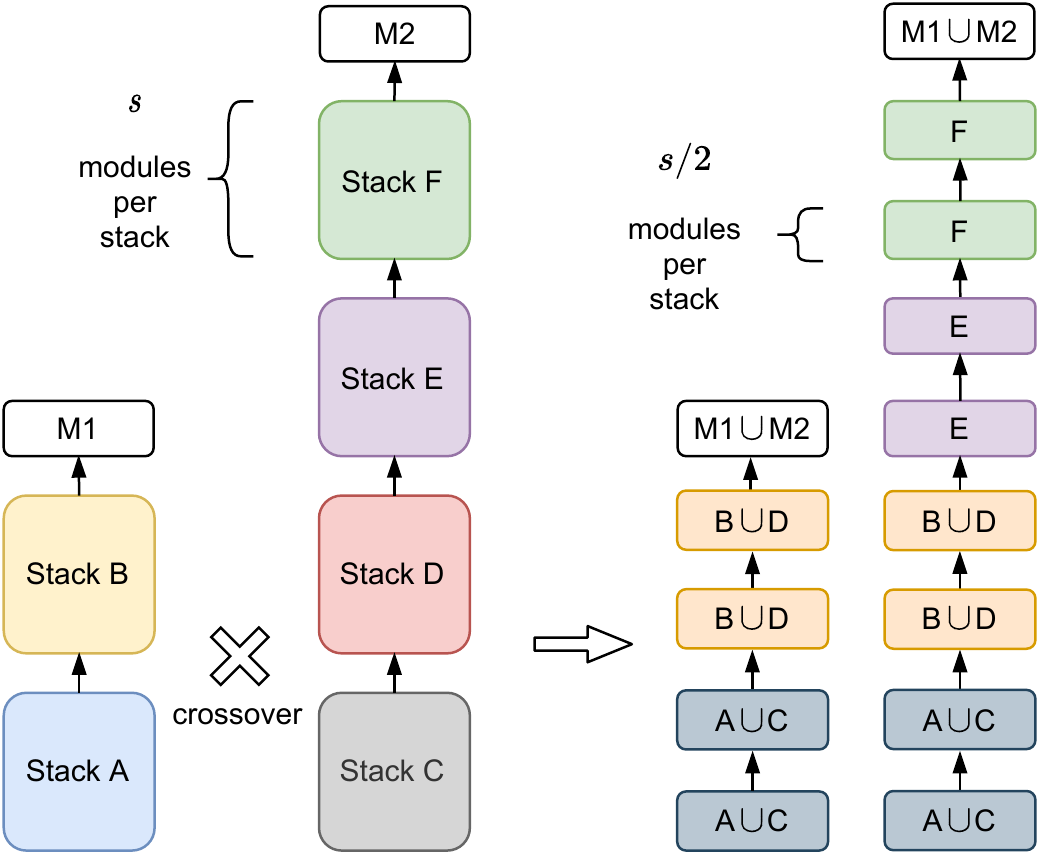}
    \caption{Crossover between neighboring models. New models are generated by creating local design spaces at 
each stack depth (where the number of modules per stack is $s$) and creating combinations of operations at half the 
stack size ($s/2$) for more granularity. MLP heads are denoted by M1 and M2, respectively.}
    \label{fig:crossover}
\end{figure}

We obtain new models in the subsequent level of the hierarchy by performing a crossover between the best models 
in the previous level (with the number of modules per stack $=s$) and their neighbors. 
Fig.~\ref{fig:crossover} presents a working example of a crossover between two neighbors. Each stack, 
respectively, has $s$ modules that are exactly the same. Once two well-performing CNNs at this level of hierarchy (or in 
other words, with a stack size of $s$) are encountered, we explore architectures that are more granular \emph{interpolants} 
between these models, i.e., with a smaller stack size, where modules can be the same only up to a smaller $s$.
For every stack depth, we create a \emph{local} space of operation blocks by considering a union of the set of 
operations used in stacks at the same depth (say, $A$ and $C$, as in Fig.~\ref{fig:crossover}). Then, new modules 
are generated by sampling from these local spaces (denoted by $A \cup C$) and stacked based on the new 
$s/K$ modules per stack ($K \in \mathbb{N}, K \le s$). As a concrete example, if stacks $A$ and $C$ had $s=10$ modules within them (again, each module in the stack being the same), then a design space from their modules is formed ($A \cup C$). Modules are then sampled from this space and stacked with a shorter stack size, say $s=5$, to generate the design space for the next level of the hierarchy. Finally, we add the modules for the MLP heads at the end based on a sample from 
the union of operations in the respective MLP heads of the neighbors. Expanding the design space in such a 
fashion retains the original hyperparameters that result in high performance while also exploring finer-grained 
internal representations learned by combinations of these hyperparameters at the same level.

\subsubsection{Isomorphism Detection}

Two computational graphs may be isomorphic. Hence, processing both would be redundant.
We use recursive hashing to detect graph isomorphism as follows~\cite{nasbench101}. For every node in the 
computational graph, we concatenate the hash of its input, the hash of that node, and its output and then take 
the hash of the result. For nodes with multiple inputs or outputs, we sort the hashes of these inputs or outputs 
and concatenate them to get a representative hash of the input or output. We use SHA256 as our hashing 
function. Doing this for all nodes and then hashing the concatenated hashes gives us the resultant 
hash of a given computational graph. We have empirically verified that this algorithm does not cause 
\textit{false positives} for the design space in consideration. Even if we had found a slight false 
positive rate, it would only have added a small amount of redundancy to the search pipeline.

\subsubsection{\texttt{CNN2vec} Embeddings}

To run architecture search on our CNN design space, we generate a library of dense embeddings (see Fig.~\ref{fig:boshnas_flowchart}(c)) for all the CNN architectures. We refer
to them as \texttt{CNN2vec} embeddings, denoted by $\mathcal{E}_C$. We do this by first calculating the Graph 
Edit Distance (GED) for all model pairs in the design space \cite{ged_networkx}. Unlike other approaches like 
the Weisfeiler-Lehman kernel, GED bakes in \emph{domain knowledge} in graph comparisons by using a weighted sum 
of node insertion, deletion, and substitution costs. For the GED computation, we first sort all possible 
operation blocks according to their computational complexity. Then, we weight the insertion and deletion cost 
for every block based on its index in this sorted list, and the substitution cost between two blocks based on the 
difference in the indices in this sorted list. GED also takes into account edge costs (set to a small value, 
$\epsilon_{edge} = 1 \times 10^{-9}$) when calculating distances between graphs. 

Some previous works have also employed deep and compute-heavy neural models for conversion of a
computational graph to an embedding~\cite{gin, d-vae, graph_encoding}. These methods were tested on small design
spaces and are not scalable to the vast spaces used in this work. For instance, D-VAE~\cite{d-vae} encodes only 
19,020 neural architectures with only 37.26\% uniqueness. GIN-based models~\cite{gin, graph_encoding} are too 
expensive to encode all models in the CNNBench design space efficiently. Hence, we use the fast and accurate 
GED-based distance metric when generating the \texttt{CNN2vec} embeddings.

Given $N$ graphs in the CNN design space of computational graphs ($\mathcal{G}$), we compute the GED for all 
possible graph pairs. This gives us a dataset of $P = {N \choose 2}$ distance pairs. To train the embedding, we 
minimize the mean-squared error as the loss function between the predicted Euclidean distance and the 
corresponding GED. For the design space in consideration, we generate embeddings of $d$ dimensions for every 
level of the hierarchy. More concretely, to train the embedding $\mathcal{E}_C$, we minimize the loss
\begin{equation*}
    \mathcal{L}_{\mathcal{E}_C} = \sum_{1 \le i \le P, 1 \le j \le P, i \ne j} \Big( \mathbf{d}(\mathcal{E}_C({g_i}), \mathcal{E}_C({g_j})) - \mathrm{GED}(g_i, g_j) \Big)^2,
\end{equation*}
where $\mathbf{d}(\cdot, \cdot)$ is the Euclidean distance and 
the $\mathrm{GED}$ is calculated for the corresponding computational graphs 
$g_i$, $g_j \in \mathcal{G}$. \texttt{CNN2vec} embeddings are superior to path encodings used in \cite{bananas} 
because they are dense and hence more amenable to search in surrogate models~\cite{nasgem}. They also decouple 
the embedding process from the search flow to speed up training~\cite{bananas}. Further, since we derive these embeddings from graph distances, we can better interpret \emph{interpolants} between two CNN architectures. Finally, since this method learns a tabular embedding by directly propagating the gradients of the above loss backward, one can train the embeddings speedily with large batch sizes and high parallelization with little memory overhead on a GPU. We use these 
embeddings in our search process.

\subsubsection{Weight Transfer among Neighboring Models}

Training each model in the design space is computationally expensive. Hence, we rely on weight sharing to 
initialize a query CNN model, thus setting the weights closer to the optimum in order to train new queries while minimizing exploration time 
(we also employ early stopping). This reduces the overall search time by 32\%~\cite{flexibert} relative to training from scratch. Previous works have implemented weight transfer between two CNN models~\cite{network_transformation, nas_lamarckian}. However, such works only support weight transfer between models with wider or deeper versions of the same computational blocks. On the other hand, CNNBench supports weight transfer between \emph{any} two models in its design space due to 
the modularity inherent in the representation and implementation. 

For every new query ($q$) that we need to train, we find $k$ nearest neighbors of its corresponding computational graph in the design space 
(we use $k=100$ in our experiments), found based on the Euclidean distance from the \texttt{CNN2vec} embeddings 
(that have a one-to-one correspondence with the GED of the respective computational graphs) of other CNNs. Here, the set of neighbors of $q$ is denoted by $N_q$, and $|N_q| = 100 \ \forall \ q$. 

Now that we have a set of 
neighbors, we need to determine which ones are more amenable to weight transfer. Naturally, we would like to 
transfer weights from the corresponding trained neighbor closest to the query, as such models 
intuitively have similar initial internal representations. We calculate this similarity using a new 
\textit{biased overlap} metric that counts the number of modules from the input to the output that are in common 
with the current graph (i.e., have exactly the same set of operations and connections). We stop counting the 
overlaps when we encounter different modules, regardless of subsequent overlaps. This ranking could lead to
more than one graph with the same \emph{biased overlap} with the current graph. Since the internal 
representations learned would depend on the subsequent set of operations as well, we break ties based on the 
embedding distance of these graphs from the current one. 

We now have a set of neighbors for every graph that are ranked based on both \emph{biased overlap} and embedding 
distance. This helps increase the probability of finding a trained neighbor with high overlap. As a hard 
constraint, we only consider transferring weights if the \emph{biased overlap fraction} ($\mathcal{O}_f(q, n) = 
\text{\emph{biased overlap}}/l_q$, where $q$ is the query model, $n \in N_q$ is the neighbor in consideration, 
and $l_q$ is the number of modules in $q$) between the queried model and its neighbor is above a threshold 
$\tau_{WT}$. If the query model meets the constraint, we transfer the weights of the shared part from the corresponding neighbor to the query and train it under this weight initialization. Otherwise, 
we may have to train the query model from scratch. We denote the weight transfer operation by $W_q \gets W_n$.

\subsubsection{BOSHNAS}
\label{sec:boshnas}


The state-of-the-art NAS technique in the CNN design space, namely BANANAS \cite{bananas}, uses an ensemble NN to 
predict uncertainty in model performance. However, ensemble NNs are dramatically more compute-heavy
than a single NN predictor. Researchers have also leveraged NNs to convert an optimization problem into a 
mixed-integer linear program (MILP) to search for better-performing points in the design space~\cite{cnma}. 
However, solving the MILP problem is computationally expensive and is often the bottleneck in this solution. 
Instead, we leverage a novel framework, namely BOSHNAS (see Fig.~\ref{fig:boshnas_flowchart}(c)), for searching over a space of CNN architectures for the first time. BOSHNAS runs gradient-based optimization using backpropagation to the input (GOBI)~\cite{cosco} on a \emph{single} and 
\emph{lightweight} NN model that predicts not only model performance, but also the epistemic and aleatoric
uncertainties. It leverages an \emph{active learning} framework to optimize the 
upper confidence bound (UCB) estimate of CNN model performance in the embedding space. We use the estimates of 
aleatoric uncertainty to further optimize the training recipe for every model in the design space. We explain
the BOSHNAS pipeline in detail next. Alg.~\ref{alg:boshnas} shows its pseudo-code.

\SetKwComment{Comment}{/* }{ */}
\begin{algorithm}[t]
\SetAlgoLined
\KwResult{\textbf{best} architecture}
 \textbf{Initialize:} overlap threshold ($\tau_{WT}$), convergence criterion, uncertainty sampling 
prob. ($\alpha_P$), diversity sampling prob. ($\beta_P$), \textbf{surrogate} model ($f$, $g$, and 
$h$) on initial corpus $\delta$, design space $g \in \mathcal{G} \Leftrightarrow x \in \Delta$; \\
 \While{convergence criterion not met}{
  wait till a worker is free\;
  \eIf{prob $\sim U(0,1) < 1 - \alpha_P - \beta_P$}{
  $\delta \gets \delta \ \cup$ \{new performance point ($x, o$)\}\; 
  fit(\textbf{surrogate}, $\delta$) using Eq.~\eqref{eq:boshnas_losses}; \\
  \label{line:fit}
    $x$ $\gets$ GOBI($f$, $h$) \Comment*{Optimization step} \label{line:opt}
    \For{$n$ in $N_x$}{
     \If{$n$ is trained \& $\mathcal{O}_f(x, n) \ge \tau$}{
      $W_x \gets W_n$\;
      send $x$ to worker\; \label{line:train}
      $\textbf{break}$\;
     }
    }
  }{
  \eIf{$1 - \alpha_P - \beta_P \le$ prob. $< 1 - \beta_P$}{
    $x$ $\gets$ $\underset{x}{\textbf{argmax}}$($k_1 \cdot \sigma + k_2 \cdot \hat{\xi}$)  \Comment*{Uncertainty sampling}
    send $x$ to worker\; \label{line:uncertainty}
  }{
    send random $x$ to worker \Comment*{Diversity sampling} \label{line:diversity}
  }
  }
 }
 \caption{BOSHNAS} 
 \label{alg:boshnas}
\end{algorithm}

\textbf{Uncertainty types:} To overcome the challenges posed by an unexplored design space, it is important to 
consider uncertainty in model predictions to guide the search process. Predicting model performance 
deterministically is not enough to estimate the next most probably best-performing model. We leverage UCB 
exploration on the predicted performance of unexplored models. This can arise from not only 
the approximations in the surrogate modeling process but also parameter initializations and variations in model 
performance due to different training recipes, namely, the \emph{epistemic} and \emph{aleatoric} uncertainties.

\textbf{Surrogate model:} BOSHNAS uses Monte-Carlo (MC) dropout \citep{mc_dropout} and a Natural Parameter 
Network (NPN) to model the epistemic and aleatoric uncertainties, respectively. The NPN not only helps with a 
distinct prediction of aleatoric uncertainty that we can use for optimizing the training recipe once we are close 
to the optimal architecture, but also serves as a superior model than Gaussian processes, Bayesian Neural 
Networks, and other Fully-Connected Neural Networks (FCNNs)~\cite{npn}. Consider the NPN network $f(x; \theta)$ 
with a CNN embedding $x$ as an input and parameters $\theta$. The output of such a network is the pair 
$(\mu, \sigma) \gets f(x; \theta)$, where $\mu$ is the predicted mean performance and $\sigma$ is the aleatoric 
uncertainty.  To model the epistemic uncertainty, we use two deep surrogate models: (1) teacher ($g$) and (2) student ($h$) networks. The teacher network is a surrogate for the performance of a CNN, using its embedding 
$x$ as an input. The teacher network is an FCNN with MC Dropout (parameters $\theta'$). To compute the epistemic 
uncertainty, we generate $n$ samples using $g(x, \theta')$. The standard deviation of the sample set is denoted 
by $\xi$. To leverage GOBI and avoid numerical gradients due to their poor performance, we use a student 
network (FCNN with parameters $\theta''$) that directly predicts the output $\hat{\xi} \gets h(x, \theta'')$, a 
surrogate of $\xi$. 

\textbf{Active learning and optimization:} For a design space $\mathcal{G}$, we first form an embedding space 
$\Delta$ by transforming all graphs in $\mathcal{G}$ using the \texttt{CNN2vec} embedding. Assuming we have the 
three networks $f, g$, and $h$ initialized on a randomly sampled set of pre-trained models ($\delta$), we use the 
following UCB estimate:
\begin{flalign}
\label{eq:ucb}
\begin{split}
    &\mathrm{UCB} = \mu + k_1 \cdot \sigma + k_2 \cdot \hat{\xi} = \Big(f(x, \theta)[0] + k_1 \cdot f(x; \theta)[1]\Big) + k_2 \cdot h(x, \theta''),
\end{split}
\end{flalign}
where $x \in \Delta$, $k_1$, and $k_2$ are hyperparameters. To generate the next CNN to test, we run GOBI using 
the AdaHessian optimizer~\citep{yao2021adahessian} that uses second-order updates to $x$ 
($\nabla^2_x \mathrm{UCB}$) up till convergence. From this, we get a new query embedding, $x'$. We find the nearest embedding for a valid CNN 
architecture based on the Euclidean distance of all available CNN architectures in the design space 
$\Delta$, giving the next closest model $x$. We then train this model (from scratch or after weight 
transfer from a nearby trained model with sufficient overlap) on the desired dataset to give the respective 
performance. Once we receive the new datapoint $(x, o)$, we train the models using the loss functions on the 
updated corpus, $\delta'$:
\begin{equation}
\label{eq:boshnas_losses}
\begin{aligned}
\mathcal{L}\textsubscript{NPN}(f, x, o) &= \sum_{(x, o) \in \delta'} \frac{(\mu - o)^2}{2 \sigma^2} + \frac{1}{2}\ln{\sigma^2},\\
\mathcal{L}\textsubscript{Teacher}(g, x, o) &= \sum_{(x, o) \in \delta'} (g(x, \theta') - o)^2,\\
\mathcal{L}\textsubscript{Student}(h, x) &= \sum_{x, \forall (x, o) \in \delta'} (h(x, \theta'') - \xi)^2,
\end{aligned}
\end{equation}
where $\mu,\sigma = f(x, \theta)$, and we obtain $\xi$ by sampling $g(x, \theta')$. The first is the aleatoric 
loss to train the NPN model \cite{npn}; the other two are squared-error loss functions. We can run multiple 
random cold restarts of GOBI to get multiple queries for the next step in the search process.

To summarize, starting from an initial pre-trained set $\delta$ in the first level of the hierarchy 
$\mathcal{G}_1$, we run until convergence the following steps in a multi-worker compute cluster. To trade off 
between exploration and exploitation, we consider two probabilities: uncertainty-based exploration ($\alpha_P$) 
and diversity-based exploration ($\beta_P$). With probability $1 - \alpha_P - \beta_P$, we run second-order GOBI using the surrogate model to maximize UCB in Eq.~\eqref{eq:ucb}. Adding the converged point 
$(x, o)$ in $\delta$, we minimize the loss values in Eq.~\eqref{eq:boshnas_losses} (line~\ref{line:fit} in 
Alg.~\ref{alg:boshnas}). We then generate a new query point, transfer weights from neighboring models, and 
train it (lines~\ref{line:opt}-\ref{line:train}). With $\alpha_P$ probability, we sample the search space using 
the combination of aleatoric and epistemic uncertainties, $k_1 \cdot \sigma + k_2 \cdot \hat{\xi}$, to find a 
point where the performance estimate is uncertain (line~\ref{line:uncertainty}). To avoid getting stuck in a 
localized search subset, we also choose a random point with probability $\beta_P$ (line~\ref{line:diversity}). 
Once we converge in the first level, we continue with subsequent levels ($\mathcal{G}_2, \mathcal{G}_3, \ldots$) by forming subsequent design spaces for each level of the hierarchy, as explained in Sections~\ref{sec:hierarchy} and \ref{sec:crossover}. The time complexity of one iteration of the active learning loop is $\mathcal{O}(n_\delta)$ where $n_\delta$ is the number of models in the corpus at that iteration.

\begin{figure}
    \centering
    \includegraphics[width=0.6\linewidth]{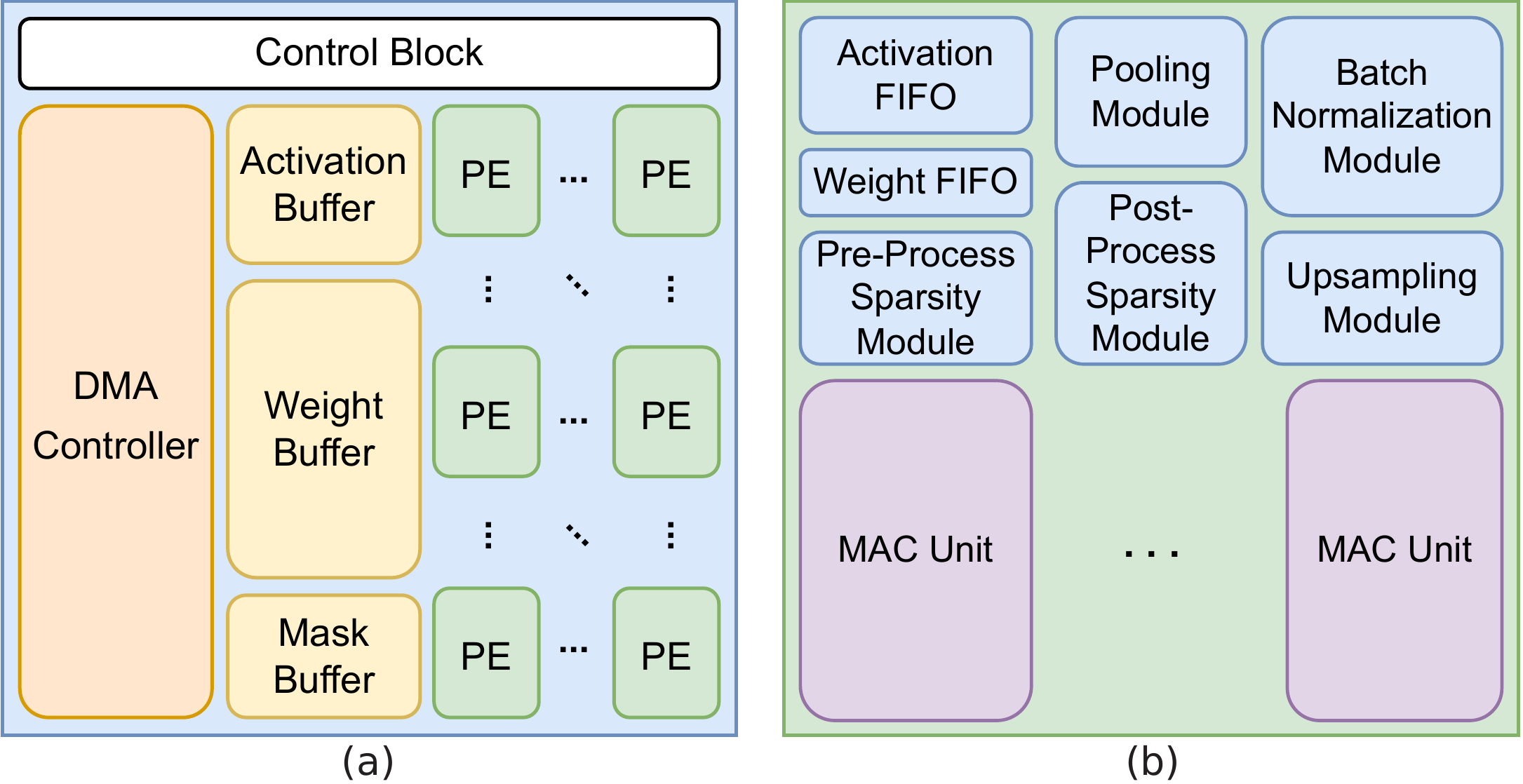}
    \caption{Accelerator/PE: (a) Generic layout of an accelerator in AccelBench. (b)~Layout of a PE in an accelerator in AccelBench.}
    \label{fig:accel_pe_layouts}
\end{figure}

\subsection{AccelBench}
\label{sec:accelbench}

Our accelerator benchmarking framework, AccelBench, is built upon a systolic 2D array architecture for generic 
ASIC-based accelerators. Fig.~\ref{fig:accel_pe_layouts}(a) shows the generic layout of the accelerators 
incorporated in our design space. Taking inspiration from SPRING~\cite{spring}, a state-of-the-art accelerator, 
all accelerators in AccelBench have a control block to handle the CNN configuration sent from the CPU and to 
control the acceleration operations. The direct memory access (DMA) controller communicates with the main memory 
system to load and store the input feature maps and filter weights from and to the on-chip buffers, respectively. We split the 
on-chip storage into three parts: \emph{activation}, \emph{weight}, and \emph{mask} buffers. The activation buffer 
stores the input feature maps and output partial sums, while the weight buffer holds the filter 
weights~\cite{spring}. We compress the data stored in both the activation and weight buffers in a zero-free 
format through a dedicated sparsity-aware binary-mask scheme to reduce the memory footprint. We design the 
mask buffer to store the binary mask vectors used in the binary-mask scheme for sparsity-aware 
acceleration (more details below). The 2D PE array executes the main computations of CNN acceleration and 
the MAC operations. AccelBench expands the design space by scaling 
various hyperparameters in a CNN accelerator, including the number of PEs, number and design of the MAC 
units, batch size used for running CNN inference, size of on-chip buffers, and size and configuration of the main 
memory system. We present the details of these scalable hyperparameters next.

\subsubsection{Processing Elements}
\label{sec:pe}

Fig.~\ref{fig:accel_pe_layouts}(b) shows the layout of the PE and its built-in modules. It buffers input activations from the feature 
maps and weight data from the filters into the activation first-in-first-out (FIFO) and 
weight FIFO pipelines, respectively. To exploit sparsity in CNN weights, in order to improve efficiency, we employ the 
binary-mask scheme used in SPRING~\cite{spring} to skip ineffectual MAC computations. This scheme uses 
two binary masks to indicate non-zero data in both input activations and filter weights. The \emph{pre-process 
sparsity module} takes in data from the FIFOs and uses the binary masks to eliminate all ineffectual MAC 
operations, e.g., when either the input activation or the weight is zero. It then sends the zero-free MAC operations to the MAC units to complete the computations. The PE passes the outputs generated from the MAC units through the \emph{post-process sparsity module} to eliminate the zeros and maintain a zero-free format to reduce 
the memory footprint. The \emph{pooling module} supports three different pooling operations, i.e., max 
pooling, average pooling, and global average pooling. The \emph{batch normalization module} executes 
batch normalization operations used in modern CNNs to reduce covariance shift~\cite{batch_norm}. Finally, the 
\emph{upsampling module} processes the upsampling operations used to upscale the feature maps (as explained in 
Section~\ref{sec:cnnbench}). 

\subsubsection{MAC Units}

\begin{figure}[t]
     \centering
     \begin{subfigure}{\linewidth}
         \centering
         \includegraphics[width=0.5\linewidth]{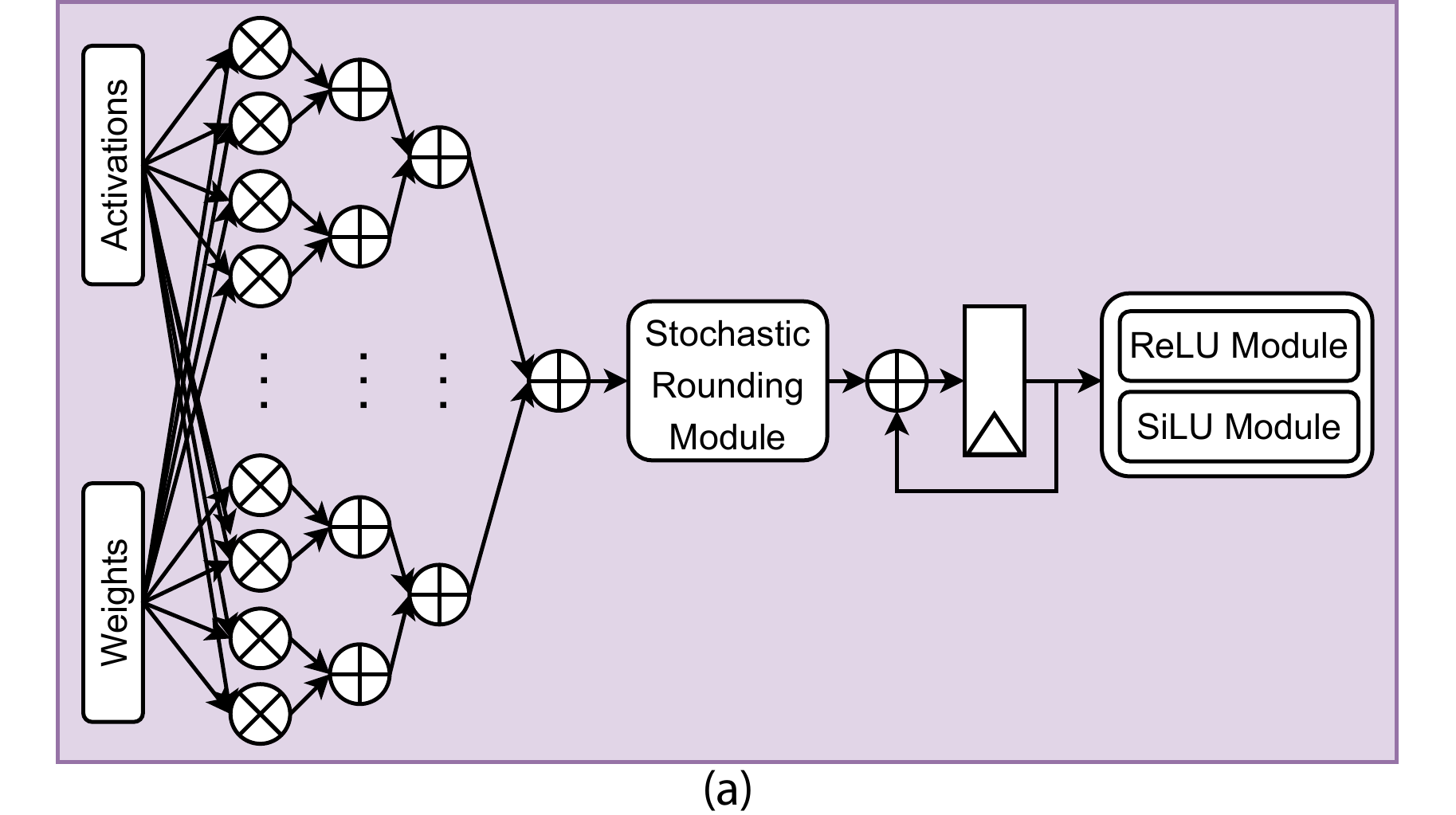}
         \label{fig:16mac}
     \end{subfigure}
     \begin{subfigure}{\linewidth}
         \centering
         \includegraphics[width=0.5\linewidth]{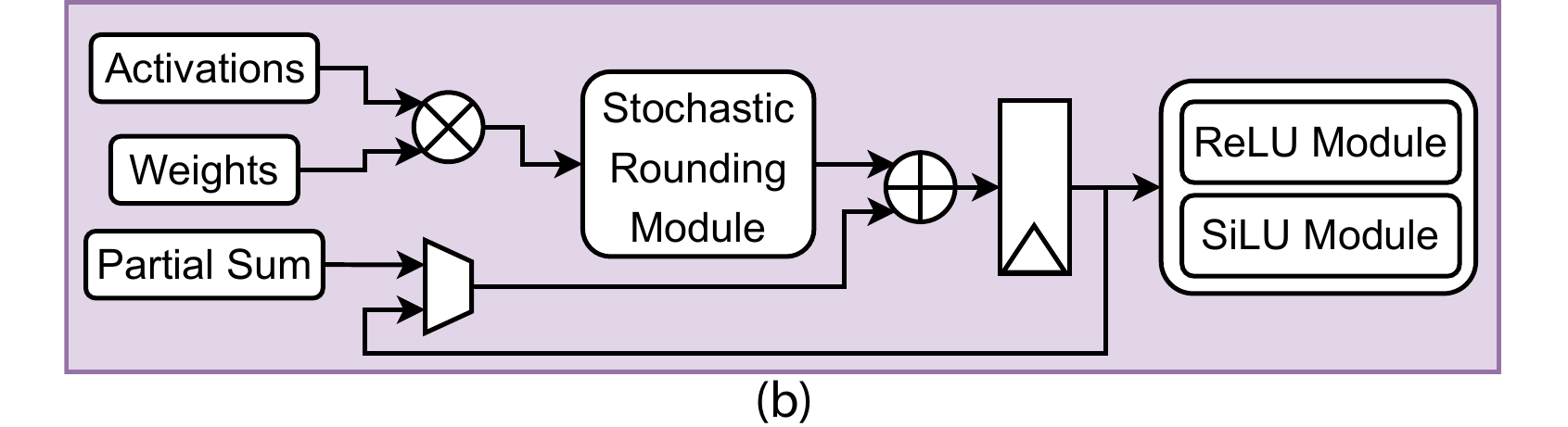}
         \label{fig:1mac}
     \end{subfigure}
        \caption{Layouts of two different MAC units: (a) 16-multiplier, (b) 1-multiplier.}
        \label{fig:mac}
\end{figure}

The MAC units, which are inside each PE, execute the MAC operations. We include two kinds of MAC units in 
AccelBench. Fig.~\ref{fig:mac}(a) shows the 
design of the 16-multiplier MAC unit. Many custom accelerators use similar designs, including DianNao, 
DaDianNao, Cambricon-X, Cambricon-S, Cnvlutin, and SPRING. The MAC unit consists of 16 multipliers that can 
perform 16 pairs of multiplications between the input activations and weights in parallel. The results from the 
multipliers feed into a 16-input adder tree to perform the accumulation operation. The accumulated sum passes 
through a dedicated \emph{stochastic rounding module} to reduce the number of bits of precision to decrease the 
memory footprint (details given later). The truncated result accumulates the previous partial sum. The module that executes the activation function: either ReLU or SiLU, takes this result as input. Fig.~\ref{fig:mac}(b) shows 
the design of the 1-multiplier MAC unit. Many custom accelerators
(including Eyeriss v1, Eyeriss v2, and ShiDianNao) use a similar design to favor certain dataflows over others. The MAC unit contains only one 
multiplier to perform one multiplication between the input activation and the filter weight. This MAC design accumulates the result with previous partial sums from the on-chip buffer or the internal register. The \emph{stochastic rounding module} again truncates this result and then sends it to either the ReLU 
or SiLU module. 

All accelerators in AccelBench use the stochastic rounding algorithm~\cite{stochastic_rounding} to quantize the data to a fixed-point 
representation. SPRING~\cite{spring} first used this algorithm. 
Unlike the traditional deterministic rounding scheme that rounds a real number to its nearest discrete integer, 
the stochastic rounding algorithm rounds a real number $x$ to $\lfloor x \rfloor$ or 
$\lfloor x \rfloor + \epsilon_R$ stochastically, as shown below:
\begin{equation}
    \label{eq:s-rounding}
    Round(x) = 
        \begin{cases}
            \lfloor x \rfloor & \text{with probability}~\frac{\lfloor x \rfloor~+~\epsilon_R~-~x}{\epsilon_R}\\
            \lfloor x \rfloor + \epsilon_R & \text{with probability}~\frac{x~-~\lfloor x \rfloor}{\epsilon_R}
        \end{cases}
\end{equation}
where $\epsilon_R$ denotes the 
smallest positive discrete integer supported in the fixed-point format and $\lfloor x \rfloor$ represents the 
largest integer multiple of $\epsilon_R$ less than or equal to $x$. The advantage of this approach over traditional 
rounding is that it does not lose information over multiple passes of the data instance. SPRING shows that using a fixed-point representation with four IL bits and 16 FL bits can 
represent a CNN with negligible accuracy loss, where IL (FL) denotes the number of bits in the integer
(fraction) part. Hence, in our experiments, we set IL = 4 and FL = 16. We use the same configuration for all hardware processing modules in AccelBench.

\subsubsection{Batch Size}

To leverage parallel computation in accelerators, AccelBench supports multiple batch sizes during
inference. More CNN operations can be computed in parallel with a larger batch size, resulting in higher 
throughput. However, processing with a large batch size requires sufficient hardware computation capacity,
e.g., sufficient PEs, MAC units, on-chip storage, and memory bandwidth. An optimal batch 
size for an accelerator would depend on the features of the targeted CNN model. AccelBench thus 
provides flexibility to maneuver through various factors simultaneously, namely hardware capacity and CNN architecture, that affect final performance; parameters like batch size impact both the hardware and CNN performance.

\subsubsection{On-chip Buffers}

To improve on-chip data bandwidth and enable sophisticated buffer design, we partition the on-chip storage into 
three parts: activation, weight, and mask buffers. The optimal on-chip buffer size depends on the throughput 
capability of the hardware computational modules and CNN model size. The proposed optimization framework finds a balance 
between area and energy while searching for an optimal CNN-accelerator pair.

\subsubsection{Main Memory}

CNN computations involve a large number of parameters, such as the input feature maps and filter weights. A 
sufficiently large main memory is crucial for storing all the CNN weights and input images (for a given batch) 
in the CNN accelerator. Moreover, a high-bandwidth interface is indispensable to keep the accelerator running at a 
high utilization rate. AccelBench supports three different main memory systems, namely, dynamic random-access 
memory (DRAM), 3D high-bandwidth memory (HBM), and monolithic 3D resistive random-access memory (RRAM). In 
addition, we provide different memory configurations for each memory type with respect to the numbers of banks, 
ranks, and channels.

To increase the off-chip bandwidth (beyond conventional DRAM) for 
CNN computations, through-silicon via (TSV) based 3D memory interfaces, such as HBM, have been used in high-end 
GPUs and specialized CNN accelerators~\cite{hbmaccel}. Unlike an HBM that uses TSVs, the monolithic 3D memory interface employs monolithic 3D integration to fabricate the chip
tier-by-tier on only one substrate wafer. The tiers are connected through monolithic inter-tier vias (MIVs), 
whose diameter is one-to-two orders of magnitude smaller than that of TSVs, enabling a much higher 
MIV density and thus a higher bandwidth~\cite{monolithic_3d_rram}. SPRING leverages this 
monolithic 3D memory system based on non-volatile RRAM to deliver high memory bandwidth and energy 
efficiency~\cite{spring}. 

\subsubsection{CNN Mapping}

\begin{figure}
    \centering
    \includegraphics[width=0.6\linewidth]{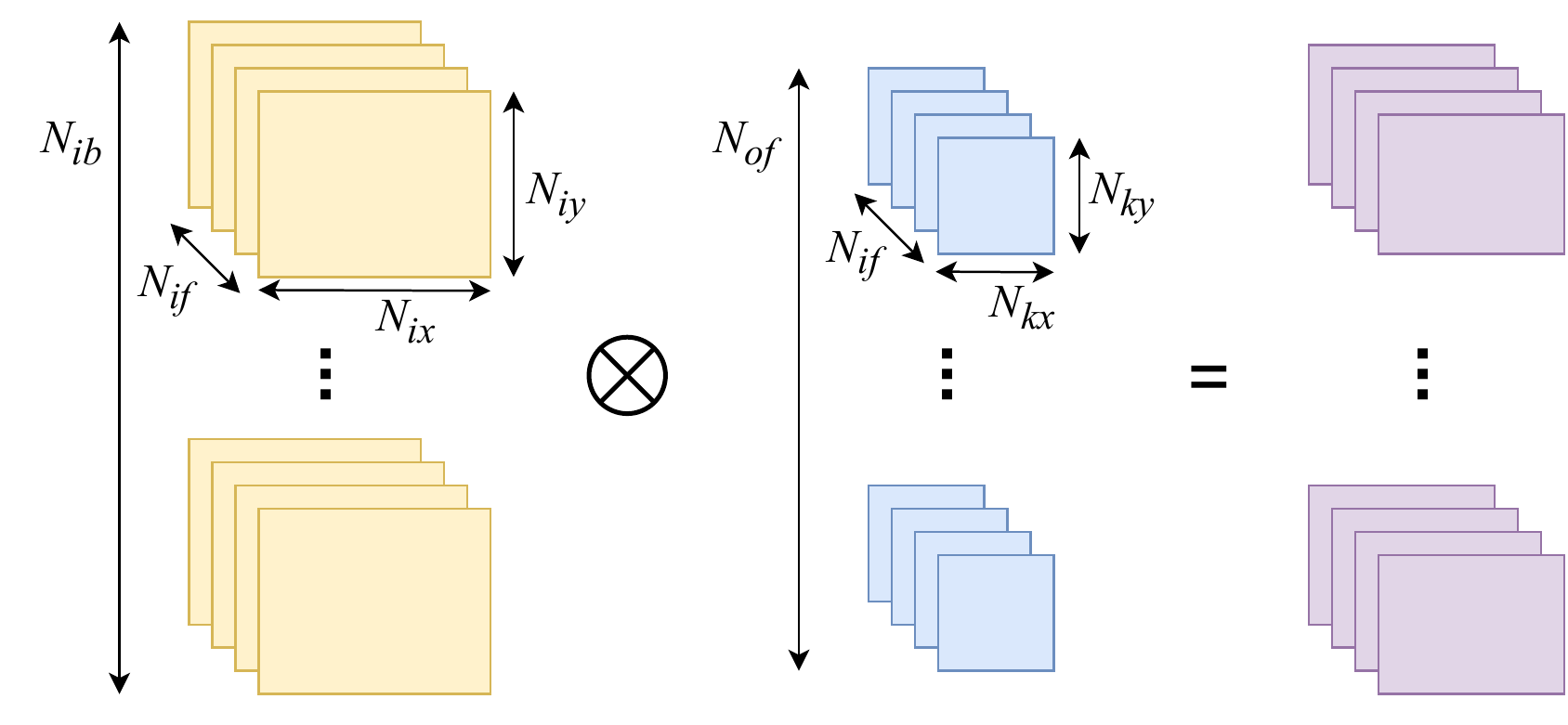}
    \caption{The convolutional layer of a CNN.}
    \label{fig:cnn}
\end{figure}

Fig.~\ref{fig:cnn} shows a convolutional layer of a CNN. To map the MAC operations of a convolutional layer onto 
an accelerator, data need to be partitioned into smaller chunks for placement onto on-chip buffers. This is 
called data \emph{tiling}~\cite{VivienneSzeBook}. After performing data tiling, one can explore the computational parallelism in a convolutional layer by unrolling the data chunks in different dimensions. This is called loop \emph{unrolling}~\cite{VivienneSzeBook}. There are a total of seven dimensions 
of parallel computations we can explore in both input feature maps and filter weights. $N_{ib}, N_{if}, N_{ix}, 
N_{iy}, N_{of}, N_{kx}$, and $N_{ky}$ denote the input batch size, number of input feature map channels, width 
and height of the input feature maps, number of output feature map channels, and width and height of the filter 
weights, respectively~\cite{softwaredefined}. We specify the number of parallel computations among these dimensions as $P_{ib}, P_{if}, 
P_{ix}, P_{iy}, P_{of}, P_{kx}$, and $P_{ky}$, respectively. We use these parameters to determine the number of 
PEs and MAC units of the accelerator. The number of PEs: $\#PEs = P_{ib} \times P_{ix} \times 
P_{iy}$. The number of MAC units inside each PE: $ \#MAC\ units = P_{of} \times P_{kx} \times P_{ky}$. We set $P_{if}$ to either 16 or 1, depending on the type of MAC unit used in the accelerator. 

Choosing the dataflow is also an important design choice while building a CNN accelerator. We choose 
the dataflow (namely OS) based on the state-of-the-art accelerator, SPRING~\cite{spring}. This accelerator selects the 
OS dataflow based on profiled performance on popular CNN architectures. Thus, OS is the dataflow of choice for all 
architectures in our design space. Some recent works have shown the advantages of reconfiguring the dataflow on a 
layer-wise basis~\cite{dataflow_opt}. However, this would add significant interconnect overhead to the architectures in 
the AccelBench design space. Altering the dataflows also requires redesigning the compiler mapping strategies, including 
the loop unrolling order and data tiling size, and the connectivity between the MAC units and PEs~\cite{VivienneSzeBook}. 
In addition, adding dataflow support would significantly expand the accelerator design space, leading to longer search 
times. The fact that hierarchical search is impossible in accelerator designs exacerbates this further. Hence, we leave 
dataflow exploration to future work.

\subsubsection{Accelerator Embeddings}

To run BOSHCODE with the AccelBench framework, we represent each accelerator with a 13-dimensional vector. The 
elements of this vector represent the hyperparameters in our design space.
We present more details in Section~\ref{sec:exp_ads}.

\subsection{Co-design Pipeline}
\label{sec:code_pipeline}

We leverage both CNNBench and AccelBench in the co-design process. We describe various parts of the co-design pipeline next.

\subsubsection{BOSHCODE}
\label{sec:boshcode}

BOSHNAS basically learns a function (comprising $f$, $g$, and $h$; c.f. Section~\ref{sec:boshnas}) that maps
the CNN embedding to model performance. It then runs GOBI to get the next CNN architecture that maximizes performance. 
BOSHCODE extends this idea to CNN-accelerator pairs. It uses the same three functions in BOSHNAS (namely 
$f$, $g$, and $h$). However, we modify these functions to incorporate design spaces of both CNNs and accelerators, 
and implement co-design. Again, we model the epistemic uncertainty by the teacher networks ($g$, with MC dropout) and a student network $h$, and the aleatoric uncertainty by an NPN ($f$). The performance measure for optimization is a convex combination of 
latency, area, dynamic energy, leakage energy, and model accuracy. Mathematically,
\begin{align}
\label{eqn:perf_metric}
\begin{split}
    \text{Performance} &= \alpha \times (1 - \text{Latency}) + \beta \times (1 - \text{Area}) + \gamma \times (1 - \text{Dynamic Energy}) \\
    &+ \delta \times (1 - \text{Leakage Energy}) + \epsilon \times \text{Accuracy}
\end{split}
\end{align}
where $\alpha + \beta + \gamma + \delta + \epsilon = 1$ are hyperparameters and we normalize the values of the individual
performance measures with their maximum values (hence, reside in the $[0,1]$ interval). Thus,
for edge applications where the power envelope of devices is highly restricted, users can set the
hyperparameters $\gamma$ and $\delta$ high. On the other hand, for server-side deployments, where accuracy
is of utmost importance, one can set $\epsilon$ high.

\subsubsection{Learning the Surrogate Function}

\begin{figure}
    \centering
    \includegraphics[width=0.5\linewidth]{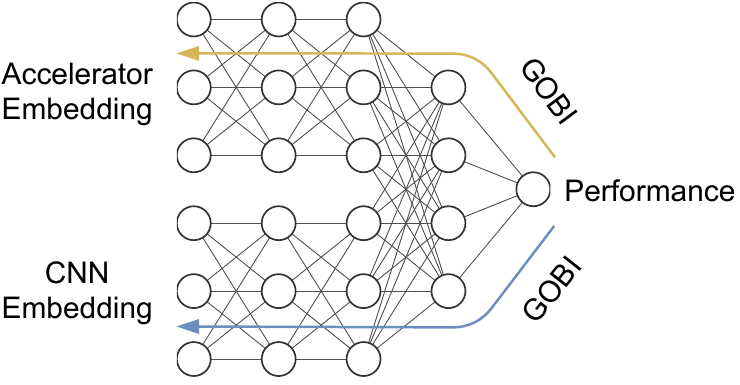}
    \caption{Schematic of the BOSHCODE teacher function. Dropout layers have been omitted for simplicity.}
    \label{fig:boshcode_nn}
\end{figure}

Using the dataset derived from mapping CNN-accelerator pairs to their respective performance values, BOSHCODE 
learns a surrogate function that predicts performance (along with the 
uncertainty) for all pairs in the design space. It leverages the active learning pipeline in BOSHNAS to query CNNs and accelerator architectures in their respective design space, to get an optimal pair that maximizes the given performance function. As noted above, BOSHCODE does not use separate models for the CNN and 
accelerator but a hybrid one instead. This aids in learning the performance of every CNN-accelerator pair, 
which is a function of the independent CNN and accelerator design decisions and their interdependence. 
In other words, we execute GOBI on representations learned on information from both the CNN and accelerator parameters and also representations learned 
specifically for either the CNN or the accelerator. 

Fig.~\ref{fig:boshcode_nn} shows a simplified schematic of the teacher network in BOSHCODE. It maps the
CNN-accelerator embeddings to the corresponding performance measures, which are a combination of the 16-dimensional \texttt{CNN2vec} embeddings and
the 13-dimensional accelerator embeddings. As explained above, the network learns separate representations
of performance for the CNN and accelerator and then combines them to give a final performance prediction. The student network learns the epistemic uncertainty from this teacher network through 
sampling (c.f. Section~\ref{sec:boshnas}). 
Then, we run GOBI on the combined and separate representations (of the student network) to find the optimal CNN-accelerator pair that maximizes the UCB estimate of the performance. Here, the gradients are backpropagated to the input (i.e., the CNN-accelerator pair) to obtain the next pair to query in the active learning loop. The algorithm iteratively evaluates CNN-accelerator pairs (using the CNNBench and AccelBench frameworks) to obtain the best-performing pair. Moreover, as explained in Section~\ref{sec:boshnas}, we implement this in a hierarchical fashion to efficiently search the massive design space of $9.3 \times 10^{820}$ CNN-accelerator pairs 
(details of design spaces given in Section~\ref{sec:exp}). More concretely, we gradually drop the stack size $s$, from 10 
to 1 over multiple iterations as the search process goes through numerous levels of the hierarchy. This is a crucial step in 
making the search feasible.

Any performance predictor has some inaccuracies. Uncertainty modeling and subsequent search mitigate the 
impact of such errors. Hence, while running the BOSHCODE pipeline, we leverage uncertainty-based sampling to achieve the 
optimal solution. When BOSHCODE reaches the optimal CNN-accelerator pair, it leverages aleatoric uncertainty in prediction 
to query the same pair multiple times. This aids in validating the optimality of the converged pair. Previous works do 
not test their surrogate model accuracy around the converged optimal point, thus leading to sub-optimal solutions. 
BOSHCODE also leverages diversity sampling to reduce modeling error on unexplored points in the design space.

\subsubsection{Inverse Design}

To add constraints to the CNN and accelerator design spaces, we limit the respective spaces to certain vectors. 
In other words, BOSHCODE only subsequently tests CNN-accelerator pairs within a pre-defined tabular design space. 
For instance, if the user wants to confine the accelerators within a certain area constraint, we restrict GOBI 
to select the nearest vector (from the reached local optimum in the continuous space) in the tabular accelerator search space
that satisfies the constraint. 

\section{Experimental Setup}
\label{sec:exp}

In this section, we discuss the setup and assumptions behind various experiments we performed in this work.

\subsection{CNN Design Space}
\label{sec:exp_cds}

As described in Section~\ref{sec:cnnbench}, we use the operation blocks of various functions, convolution, 
pooling, MLP layers, activation functions, etc., to define the design space. This results in 618 operation 
blocks in all. We describe these operations next.
\begin{itemize}
    \item Channel shuffle in groups of \{1, 2, 4, 8\}~\cite{shufflenet}.
    \item Dropout with probability of \{0.1, 0.11, 0.2, 0.3, \ldots, 0.9\}~\cite{dropout}.
    \item Upsampling to size of \{240, 260, 300, 380, 465, 528, 600, 800\}~\cite{efficientnet}.
    \item Maxpool and Avgpool in kernel sizes of \{$3 \times 3$, $5 \times 5$\} with a padding of either 0 or 1 and stride 
of 1 or 2.
    \item Convolution in kernel sizes of \{$1 \times 1$, $3 \times 3$, $5 \times 5$, $7 \times 7$, $11 \times 11$\} with output channels in 
\{4,$\ldots$, 8256\} (98 values),
with groups in \{4, 8, 16\} or depth-wise layers where the groups are equal to the number of input channels. The
padding and strides supported are from \{0, 1, 2, 3\} and \{1, 2, 4\}, respectively. For all convolutional layers, we support both 
ReLU and SiLU activations~\cite{silu}.
\item Flatten and global-average-pool for flattening the output from the convolutional layers. Global-average-pool computes an average across the spatial dimensions to output a vector of length equal to the number of output channels from the convolutional layers~\cite{global_avg_pool}.
    \item MLP layers with the number of hidden neurons in \{84, 120, 1024, 4096\} and finally the number of classes.
\end{itemize}

To limit the size of the design space, we do not consider all combinations but only those prevalent 
in popular CNN architectures. We limit the CNN computational graphs to a depth of 90 modules (excluding the final head module), each 
module with a maximum of five vertices (including the \emph{input} and \emph{output} operations) and limited to eight edges in modules with convolutional operations. For the final head 
module, we limit the number of vertices to eight (it only has linear feed-forward connections of fully-connected 
and dropout layers). This 
leads to a total design space of size $4.239 \times 10^{812}$. This is much larger than any other NAS design space studied 
before~\cite{nasbench101, nasbench301, once_for_all}. We form the first level of the hierarchy by creating graphs with a stack size of 10, resulting in $2.089 \times 10^{85}$ CNN architectures in the first level (details in Section~\ref{sec:hierarchy}). This not only eases the search for the best-performing CNN but also reduces the disparity in CNN and accelerator search space sizes that might lead to challenges in optimization.

The \texttt{CNN2vec} embedding has dimension set to $d=16$ after running grid search. To do this, we 
minimize the distance prediction error while also keeping $d$ small using knee-point detection. We obtain the hyperparameter 
values for BOSHNAS  through grid search. We use overlap 
threshold $\tau_{WT}=80\%$ and constants $k_1 = k_2 = 0.5$ (see Section~\ref{sec:cnnbench}) in our experiments. 
We set the uncertainty and diversity sampling probabilities to $\alpha_P = 0.1$, $\beta_P = 0.1$. The algorithm reaches the convergence 
criterion when the change in performance is within $10^{-4}$ after five iterations.

We set a fixed training time of 200 epochs for every CNN on the CIFAR-10 dataset~\cite{cifar10}. Early stopping can result in lower training time for certain CNNs. We use a batch size 
of 128 for training. We automatically tune the training recipe for every CNN to tune the 
hyperparameter values. We leverage the Asynchronous Successive Halving scheduler \cite{asha}, 
implemented using random search over different hyperparameter values, in the training recipe. 
CNNBench supports various optimizers including Adam, 
AdamW~\cite{adamw}, etc., along with various learning-rate schedulers including cosine annealing, exponential, etc. We sample learning rates in a log-uniform manner in the 
$[1\times 10^{-5}, 1 \times 10^{-2}]$ interval. We sample other optimizer parameters as follows: 
$\beta_1 \sim \mathcal{U}(0.8, 0.95)$, $\beta_2 \sim \mathcal{U}(0.9, 0.999)$, and weight decay 
$\ln(\lambda) \sim \mathcal{U}(\ln(1 \times 10^{-5}), \ln(1 \times 10^{-3}))$. Here, $\mathcal{U}$ refers to the uniform distribution. 

\subsection{Accelerator Design Space}
\label{sec:exp_ads}

\begin{table}[t]
\centering
\caption{Hyperparameter values used in the proposed accelerator design space.}
\resizebox{0.6\columnwidth}{!}{
\begin{tabular}{@{}ll@{}} 
    \toprule
    \textbf{Hyperparameter} & \textbf{Permissible values} \\
    \midrule
    $P_{ib}$ & 1, 2, 4 \\ 
    $P_{if}$ & 1, 16\\ 
    $P_{ix}$ & 1 $\sim$ 8 \\ 
    $P_{iy}$ & 1 $\sim$ 8 \\
    $P_{of}$ & 1, 2, 4, 8 \\
    $P_{kx}$ and $P_{ky}$ & 1, 3, 5, 7 \\
    \emph{batch size} & 1, 64, 128, 256, 512 \\
    \emph{activation buffer size} & 1MB $\sim$ 24MB, in multiples of 2\\
    \emph{weight buffer size} & 1MB $\sim$ 24MB, in multiples of 2\\
    \emph{mask buffer size} & 1MB, 2MB, 3MB, 4MB\\
    \multirow{3}{*}{\emph{main memory type}} & 1: Monolithic 3D RRAM \\
     & 2: Off-chip DRAM \\
     & 3: HBM \\
    \multirow{5}{*}{\emph{main memory configuration}} &  RRAM: 1: [16, 2, 2], 2: [8, 2, 4], 3: [4, 2, 8], \\
     & 4: [2, 2, 16], 5: [32, 2, 1], 6: [1, 2, 32] \\
     & DRAM: 1: [16, 2, 2], 2: [8, 2, 4], 3: [32, 2, 1], \\
     & 4: [16, 4, 1] \\
     & HBM: 1: [32, 1, 4] \\
     \bottomrule
\end{tabular}}
\label{tbl:parameters}
\end{table}

Fig.~\ref{fig:code_pipeline}(a) shows the workflow of AccelBench. We implement all the supported modules in 
AccelBench at the register-transfer level (RTL) with System Verilog. We synthesize the RTL modules with 
Synopsys Design Compiler~\cite{dc} using a 14nm FinFET cell library~\cite{14nm} to estimate delay, power consumption, and area. We 
model on-chip buffers with FinCACTI~\cite{fincacti} and the main memory systems with NVMain~\cite{nvsim, nvmain}. Finally, we plug in the estimated delay, power 
consumption, and area of the RTL modules, on-chip buffers, and main memory systems into a custom cycle-accurate 
accelerator simulator implemented in Python. AccelBench's accelerator simulator takes the Python object of the 
CNN model generated using CNNBench as its input and estimates the computation latency, energy consumption, and
area of the accelerator in consideration. 

As mentioned in Section~\ref{sec:accelbench}, we encode each accelerator in AccelBench with a 13-dimensional 
vector, which contains the hyperparameter values used to expand the accelerator design space. We show the ranges of all hyperparameter values in Table~\ref{tbl:parameters}. Note that we set $P_{kx} = P_{ky}$ to match the conventional square matrix structure of the filter weights. The three entries in every list for main memory configurations are the number of banks, ranks, and channels, respectively. Using these respective ranges, we get a design space with 
$2.28 \times 10^8$ accelerators, much larger than in any previous work~\cite{hw_sw_co-exp, bobw, nahas, naas}. 

\subsection{Co-design and Baselines}
\label{sec:exp_baselines}

For running BOSHCODE, we use the following parameter values to obtain the performance measure: $\alpha = 0.2$, $\beta
= 0.1$, $\gamma = 0.2$, $\delta = 0.2$, $\epsilon = 0.3$ (see Section~\ref{sec:code_pipeline}). Other hyperparameters
are the same as described in Section~\ref{sec:exp_cds} 
All models are trained on NVIDIA A100 GPUs and 2.6 GHz AMD EPYC Rome processors. 
The entire process of training BOSHCODE to get the best CNN-accelerator pair took around 600 GPU-days.

As described in Table~\ref{tbl:accelerators}, we incorporate various accelerators into our AccelBench framework.
For a fair comparison of various performance measures, we implement an analogous accelerator in the AccelBench design
space by transferring the design decisions into a 13-dimensional vector. However, direct comparison with respective results is
challenging since each work uses a separate baseline for comparison and the reported results are often normalized.
Further, individual works implement accelerators with different hardware modules, CNN mapping strategies, technology nodes, and clock frequencies. Thus, AccelBench serves as a common benchmarking platform that compares diverse accelerator architectures 
along fairly-determined parameter values and performance measures.

We also compare our work with various co-design baselines. Co-Exploration~\cite{hw_sw_co-exp} explores the hardware
and software spaces simultaneously, with different objectives: optimizing either the hardware design decisions (OptHW)
or the software CNN architecture (OptSW). It works with up to three Xilinx FPGAs (XC7Z015) and searches for a pipelining strategy. BoBW~\cite{bobw} implements RL-based co-design on the NASBench-101
dataset~\cite{nasbench101} for CNNs with a library of FPGAs. It uses CHaiDNN, a library for acceleration of CNNs on FPGAs~\cite{chaidnn}. The CHaiDNN FPGA accelerator has various configurable parameters~\cite{bobw}. NASAIC~\cite{nasaic} and NAAS~\cite{naas} leverage RL and
ES, respectively, on a space of accelerators based on the NVDLA~\cite{nvdla},
ShiDianNao~\cite{shidiannao}, etc., dataflow templates. We also add a state-of-the-art differentiable search
based baseline, i.e., Auto-NBA~\cite{auto_nba}, which has a reasonably large design space of size $5.3 \times 10^{63}$. It employs the Accelergy platform, an energy estimator tool for accelerators~\cite{accelergy}. Its design space is based on recent FPGA-based accelerators~\cite{max_cnn, dnnbuilder} and adopts a chunk-wise pipelined architecture.

\section{Results}
\label{sec:res}

In this section, we validate our claims by comparing the CODEBench framework with the aforementioned baselines.

\subsection{BOSHNAS on a CNN Design Space}

\begin{figure}[t]
    \centering
    \includegraphics[width=0.6\linewidth]{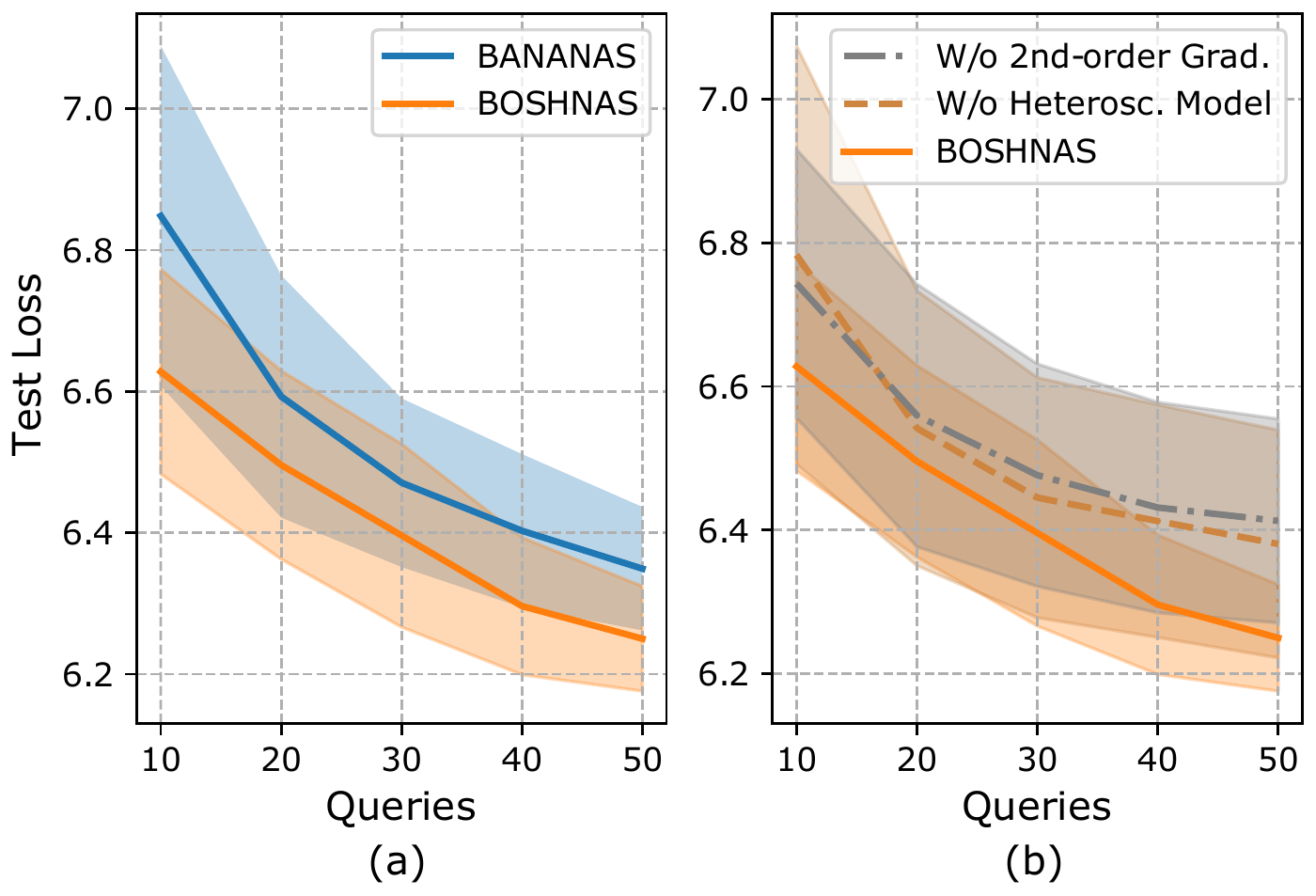}
    \caption{Performance of BOSHNAS and BANANAS: (a) NASBench-101 dataset and (b) ablation analysis of BOSHNAS. 50 trials are 
run for all techniques. Plotted with 90\% confidence intervals.}
    \label{fig:boshnas_vs_baselines_and_ablation}
\end{figure}

Fig.~\ref{fig:boshnas_vs_baselines_and_ablation}(a) shows the performance of BOSHNAS with respect to the 
state-of-the-art NAS technique, BANANAS~\cite{bananas}, on the NASBench-101 dataset~\cite{nasbench101}. The figure plots the final loss on the test set for CNNs found after a given number of queries from either NAS technique. As we can see, BOSHNAS 
results in a lower average test loss than BANANAS after 50 queries on the NASBench-101 design space. 
Fig.~\ref{fig:boshnas_vs_baselines_and_ablation}(b) shows an ablation analysis of BOSHNAS on the NASBench-101 
dataset, with the BOSHNAS model being used once without second-order gradients and once without modeling heteroscedasticity, i.e., the NPN model $f$. 
These plots justify the need for heteroscedastic modeling and second-order gradients in the space of CNN models. The heteroscedastic model forces
the optimization of the training recipe when the framework approaches optimal architectural design decisions. Second-order gradients, on the other hand, help the search avoid local optima and saddle points, and also aid faster convergence.

\subsection{Co-design vs. One-sided Approaches}

\begin{figure}[t]
    \centering
    \includegraphics[width=0.5\linewidth]{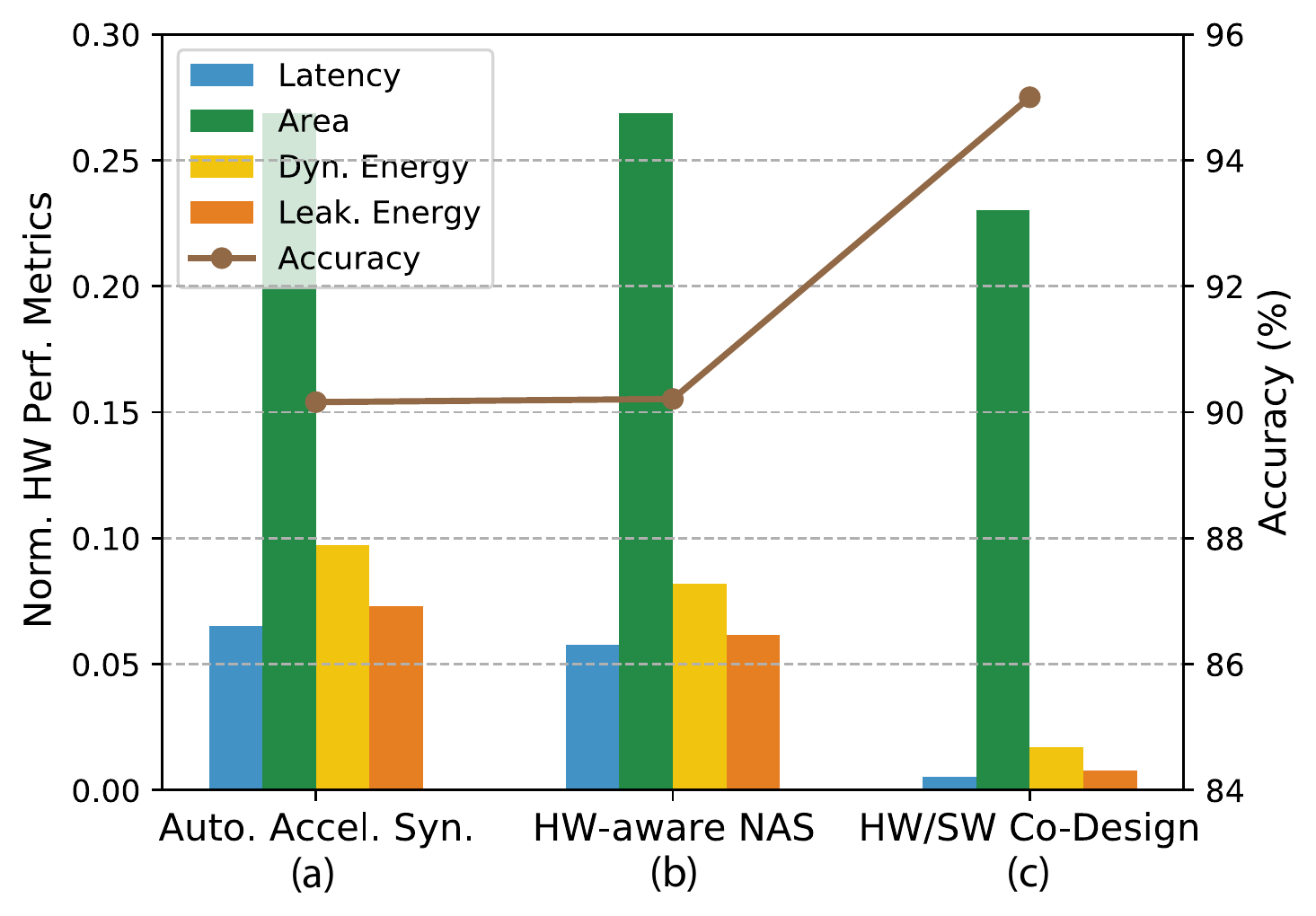}
    \caption{Performance measures when three approaches are applied to the design space: (a) fixing CNN to MobileNet-V2 and 
searching the accelerator space, (b) fixing the accelerator to SPRING~\cite{spring} and performing NAS, and (c) co-design 
in the space of CNN-accelerator pairs.}
    \label{fig:code_vs_one-sided_approaches}
\end{figure} 

Fig.~\ref{fig:code_vs_one-sided_approaches} highlights the benefits of hardware-software co-design over one-sided approaches, 
namely automatic accelerator synthesis and hardware-aware NAS, as described in Section~\ref{sec:back_rel}. We formulate the CNN and
accelerator design spaces as per the hyperparameter ranges presented in Sections~\ref{sec:exp_cds} and
\ref{sec:exp_ads}. Then, we search for a CNN-accelerator pair that optimizes the selected performance measure, which is a combination of
latency, area, dynamic and leakage energies, and accuracy (see Eq.~\eqref{eqn:perf_metric}) for every CNN-accelerator
pair. 

Fig.~\ref{fig:code_vs_one-sided_approaches}(a) shows the best normalized performance measures achieved after searching
the space of accelerator design decisions using BOSHCODE. Here, we force gradients to the CNN embedding
to zero to find the next accelerator to simulate in the search process. Fig.~\ref{fig:code_vs_one-sided_approaches}(b) shows the results when we explore the CNN design space while 
fixing the accelerator instead. In this case, we force the gradients to the accelerator embedding to zero in 
the two-input BOSHCODE network. Fixing the accelerator and searching the CNN design space does not result in noticeable
gains in accuracy but results in slight improvements in latency and energy values. This is due to the search 
process landing on another equally performing CNN, but one which is smaller in terms of the number of parameters (reduced from 
3.4M to 2.9M). This also improves hardware resource utilization. Finally,
Fig.~\ref{fig:code_vs_one-sided_approaches}(c) shows the best performance achieved from co-design in the
CNN and accelerator spaces using our proposed BOSHCODE approach. This enables higher flexibility in search, resulting in 
much higher accuracy and significant improvements in the hardware performance measures.
Co-design achieves \accuracyIncreaseCvA higher model accuracy on the CIFAR-10
dataset, while enabling \latencyReductionCvA lower latency (averaged per batch of
images), \energyReductionCvA lower total energy consumption (including both dynamic and
leakage energies), and \areaReductionCvA lower chip area, when compared to automatic
accelerator synthesis. On the other hand, relative to hardware-aware NAS, co-design achieves \accuracyIncreaseCvB higher model accuracy at \latencyReductionCvB lower latency, and \areaReductionCvB lower chip area. Fig.~\ref{fig:code_vs_one-sided_approaches} uses the following
values for normalization: 9ms for latency, 774mm$^2$ for chip area, 735mJ for dynamic energy, and 280mJ for leakage energy. 

\begin{figure*}[t]
    \centering
    \includegraphics[width=\textwidth]{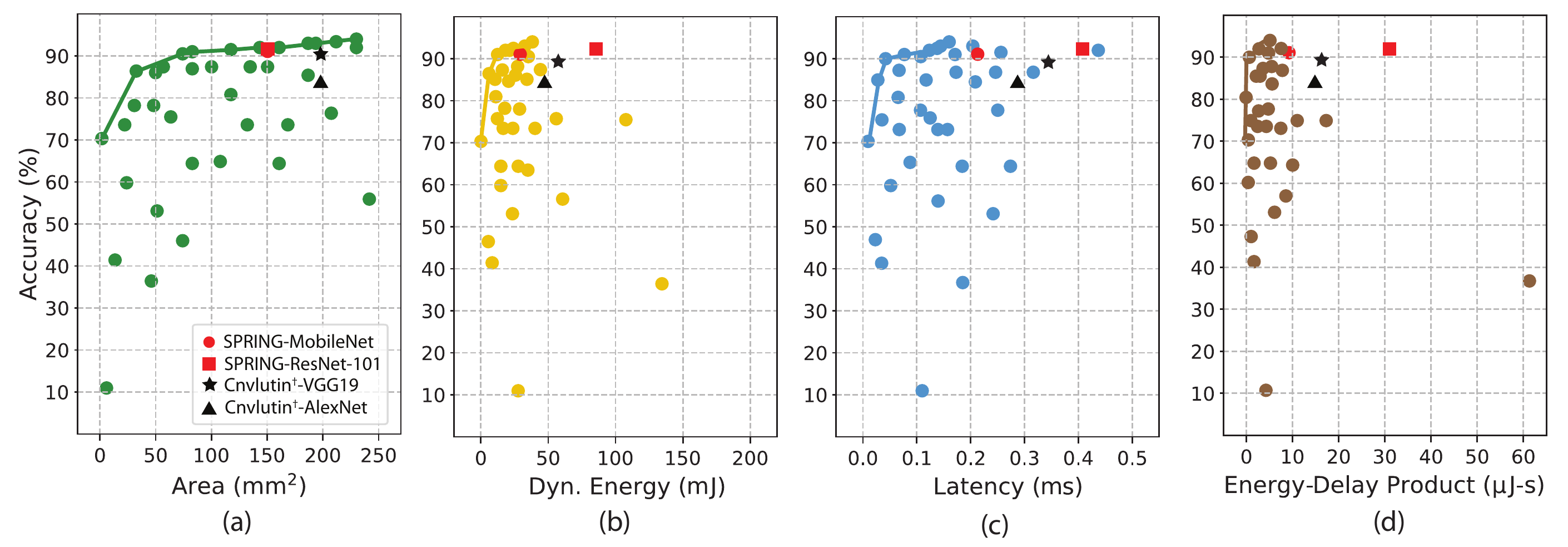}
    \caption{Pareto frontier between accuracy and (a) chip area, (b) dynamic energy consumption, (c) latency, and (d)
energy-delay product (calculated using the sum of dynamic and leakage energies). Reduced number of points (compared to 
245 trained) have been plotted for clarity. Points far from the Pareto frontier have been neglected in the plot. 
$\ssymbol{2}$Cnvlutin-like~\cite{cnvlutin} accelerator (used in conjunction with
VGG19 and AlexNet) implemented in AccelBench for fair comparisons 
(see Section~\ref{sec:exp_baselines}).}
    \label{fig:pareto_frontiers}
\end{figure*}

\subsection{Optimal CNN-accelerator Pair}

Fig.~\ref{fig:pareto_frontiers} shows the Pareto frontiers of CNN-accelerator pairs in our design space. We have also 
included other state-of-the-art pairs proposed in the literature, including MobileNet~\cite{mobilenetv2} 
and ResNet-101 \cite{resnet} on SPRING~\cite{spring} and VGG-19 \cite{vgg} and AlexNet \cite{alexnet} on Cnvlutin~\cite{cnvlutin}. One can observe these existing CNN-accelerator pairs to be far from the Pareto frontier. In other words, BOSHCODE finds pairs that 
outperform previously proposed CNN architectures and accelerator designs in user-defined performance measures. 

\begin{table}[t!]
\caption{Comparison of optimal CNN-accelerator pair with a state-of-the-art. Top1 accuracy is reported for the ImageNet dataset.}
\resizebox{0.6\columnwidth}{!}{
\begin{tabular}{@{}l|cc|cc@{}}
\toprule
Performance measure & \multicolumn{2}{c|}{CIFAR-10} & \multicolumn{2}{c}{ImageNet} \\ \midrule
 & S-MobileNet & S*-ResNet* & S-MobileNet & S*-ResNet* \\ \midrule
Latency (ms) & 0.22 & \textbf{0.09} & 0.57 & \textbf{0.32} \\
Area (mm$^2$) & \textbf{152} & 178 & \textbf{152} & 178 \\
Dynamic Energy (mJ) & 28.4 & \textbf{14.7} & 41.2 & \textbf{37.5} \\
Leakage Energy (mJ) & 19.8 & \textbf{4.2} & 6.3 & \textbf{4.7} \\
Accuracy (\%) & 93.4 & \textbf{94.8} & 72.1 & \textbf{75.8} \\ \bottomrule
\end{tabular}}
\label{tbl:opt_pair}
\end{table}

For the performance measure defined by the convex combination of different parameter values (namely, latency, area, dynamic
energy, leakage energy, and model accuracy, as calculated using Eq.~\eqref{eqn:perf_metric}), we get an optimal 
CNN-accelerator pair by running BOSHCODE on the entire design space. The optimal CNN is close to ResNet-50 (we call it
ResNet$^*$) and the optimal accelerator is similar to SPRING (we call it S$^*$). Table~\ref{tbl:opt_pair}
compares this optimal pair (denoted by S$^*$-ResNet$^*$) with the state-of-the-art baseline in various performance measures (we choose MobileNet
with SPRING as it gives the highest performance measure compared to other off-the-shelf CNNs, and denote the pair by S-MobileNet). BOSHCODE achieves \latencyReduction lower latency, \energyReduction lower energy, and \accuracyIncrease higher
accuracy compared to the state-of-the-art CNN-accelerator pair at the cost of \areaIncrease increase in area. A different 
convex combination of hyperparameter values ($\alpha, \beta, \gamma, \delta,$ and $\epsilon$) would result in BOSHCODE 
converging to a different CNN-accelerator pair. In our experiments, a low value of $\beta = 0.1$, which corresponds to the 
weight contributed to the chip area, results in a slightly higher area relative to the state-of-the-art pair. When comparing
the two pairs on the ImageNet dataset~\cite{imagenet}, we see a \latencyReductionImageNet~lower latency,
\energyReductionImageNet~lower energy consumption, and \accuracyIncreaseImageNet higher model accuracy. Due to the 
transferability of the optimal model from our search space to the ImageNet dataset, our optimal pair also outperforms other 
state-of-the-art approaches like differentiable search that are directly trained on the ImageNet dataset. For context, 
ProxylessNAS~\cite{proxyless_nas} achieves 74.6\% Top1 accuracy with a latency of 
78ms on a Google Pixel 1 smartphone. The accuracy is 1.2\% lower than that of S$^*$-ResNet$^*$ and the latency is 243.75$\times$ higher than that of S$^*$-ResNet$^*$.

\subsection{Co-design with an Expanded Design Space}

\begin{table*}[t]
\caption{Comparison of baseline and proposed co-design techniques. Energy values reported are for each input
frame.}
\centering
\resizebox{\columnwidth}{!}{
\begin{tabular}{l@{\hskip 0.2in}|@{\hskip 0.2in}c@{\hskip 0.2in}c@{\hskip 0.2in}c@{\hskip 0.2in}c@{\hskip 0.2in}c@{\hskip 0.2in}c@{\hskip 0.2in}c}
\toprule
Framework & Platform & Search space & Accuracy (\%) & Area (mm$^2$) & FPS (s$^{-1}$) & EDP ($\mu$J-s) \\
\midrule
\multicolumn{7}{c}{\textbf{Baselines}} \\ 
\midrule
Co-Exploration (OptHW)~\cite{hw_sw_co-exp} & FPGA & 5.9 $\times$ 10$^3$ & 80.2 & - & 130 & 1612.9 \\
Co-Exploration (OptSW)~\cite{hw_sw_co-exp} & FPGA & 5.9 $\times$ 10$^3$ & 85.2 & - & 130 & 1612.9 \\
BoBW~\cite{bobw} & FPGA & 4 $\times$ 10$^9$ & 94.2 & \textbf{102} & 8.1 & - \\ \midrule
NASAIC~\cite{nasaic} & DLA & 1.1 $\times$ 10$^6$ & 93.2 & 525 & 1749 & 571.8 \\
NAAS~\cite{naas} & DLA & 1 $\times$ 10$^{24}$ & 93.2 & 525 & 6562 & 365.7 \\
\midrule
Auto-NBA~\cite{auto_nba} & Accelergy~\cite{accelergy} & 5.3 $\times$ 10$^{63}$ & 93.3 & 710 & 320 & 18.7 \\
\midrule
\multicolumn{7}{c}{\textbf{Ablation Analysis}} \\ 
\midrule
\textbf{Hardware-Aware NAS (Ours)} & AccelBench & 4.2 $\times$ 10$^{812}$ & 91.8 & 245 & 1709 & 45.8 \\
\textbf{CODEBench (Ours; DRAM only)} & AccelBench & 1.6 $\times$ 10$^{820}$ & 93.9 & 184 & 8620 & 14.8 \\ 
\textbf{CODEBench (Ours)} & AccelBench & 9.3 $\times$ 10$^{820}$ & \textbf{94.8} & 178 & \textbf{11,111} & \textbf{1.7} \\ \bottomrule
\end{tabular}}
\label{tbl:boshcode_vs_baselines}
\end{table*}

Table~\ref{tbl:boshcode_vs_baselines} presents a detailed comparison of results for the CODEBench framework and various co-design baselines. Here, Co-Exploration~\cite{hw_sw_co-exp} runs co-design in a design space of three Xilinx FPGAs
(XC7Z015; fabricated on 28nm process and normalized to 14nm), chip area for which is unknown. BoBW~\cite{bobw} does not model energy consumption of the
hardware architectures in its design space. For NASAIC~\cite{nasaic} and NAAS~\cite{naas}, we report the die area for the Xavier system-on-chip that has two 
instances of NVDLA~\cite{nvdla}, fabricated in a 12nm process~\cite{xavier_soc} and normalized to 14nm~\cite{technology_norm}. For NASAIC and NAAS, we assume a 700MHz clock-rate assumed for calculation of FPS from the reported latency in cycles~\cite{naas}. For Auto-NBA, we use Auto-NBA-Mixed~\cite{auto_nba} to report the value of FPS. Moreover, the same baseline reported the EDP in J-cycles and we assumed a 700MHz clock-rate for conversion to $\mu$J-s.

Our framework not only achieves the highest model accuracy on the CIFAR-10 dataset but also improves upon other hardware 
performance measures including frame-rate per second (FPS) or throughput, and EDP. CODEBench 
achieves \accuracyIncreaseANBA higher accuracy and \fpsIncreaseANBA higher throughput while having \edpReductionANBA 
lower EDP and requiring \areaReductionANBA lower chip area, when compared to a state-of-the-art co-design framework, i.e., 
Auto-NBA.
These substantial gains compared to traditional co-design frameworks became possible due to a search over a massive design 
space of CNN-accelerator pairs (9.3 $\times$ 10$^{820}$), compared to only 5.3 $\times$ 10$^{63}$ in Auto-NBA. We can attribute 
high reductions in energy consumption to the use of sparsity-aware computation, monolithic 3D RRAM interface 
instead of an off-chip DRAM~\cite{monolithic_3d_rram}, etc. For fair comparisons, we have normalized the results 
for different technology nodes~\cite{technology_norm} and added a CODEBench optimal pair with the constraint that RRAM is 
unavailable. The co-design case outperforms hardware-aware NAS with a lower chip area and a higher throughput. We attribute this to a smaller CNN model that is a better fit for the searched accelerator, resulting in better utilization of 
resources and parallelism.

\subsection{Discussion}

For running the co-design pipeline, we trained all CNN models in our design space on the CIFAR-10 dataset. As seen from Table~\ref{tbl:opt_pair}, and also noted above, performance of models on CIFAR-10 directly 
translates to that on ImageNet as well~\cite{transferable_learning}. Running the BOSHCODE framework on the 
ImageNet dataset would incur high compute costs. Hence, this trend of transferable performance is of utmost importance to
scalable search~\cite{nasbench101}. Due to the incorporation of highly diverse CNNs and accelerator architectures, future 
researchers can directly use these surrogate models to minimize time and costs, and quickly search specialized subset spaces of 
interest.
Additionally, we could further expand both the CNN and the accelerator design spaces with more operations in the former and more 
hardware modules and memory types in the latter. In this context, we could also employ the co-design pipeline for in-memory 
accelerators and neuromorphic architectures~\cite{gibbon, neuromorphic}. Due to the proposed scalable method, one can search even larger CNN design spaces with a more aggressive hierarchical search, i.e., increasing granularity in fewer steps. One could also exploit dynamic CNN model inference~\cite{dynamic_inf} and on-the-fly quantization during inference~\cite{pact} to further reduce 
latency and energy consumption. We leave these extensions to future work.

\section{Conclusion}
\label{sec:conc}

In this work, we presented CODEBench, a unified benchmarking framework for simulating and modeling CNN-accelerator pairs and their 
performance measures (model accuracy, latency, area, dynamic energy, and leakage energy). We developed a novel 
CNN benchmarking framework, CNNBench, that leverages the proposed \texttt{CNN2vec} embedding scheme and BOSHNAS 
(that uses this embedding scheme, Bayesian modeling, and second-order optimization) to efficiently search for 
better-performing CNN models within a given design space. Our proposed CNN design space is richer compared
to that in any previous work. In order to optimize hardware performance at the same time, we also proposed a 
new accelerator benchmarking pipeline, AccelBench. It targets not only a vast design space of ASIC-based accelerator architectures but also simulates various performance measures of these hardware designs. We then 
use a novel co-design approach, namely BOSHCODE, that searches for the best CNN-accelerator pair. Our proposed 
approach results in a CNN-accelerator pair with \accuracyIncrease higher CNN accuracy compared to a state-of-the-art 
pair, i.e., SPRING-MobileNet, while enabling \latencyReduction lower latency and \energyReduction lower energy 
values. CODEBench outperforms the state-of-the-art co-design framework, i.e., Auto-NBA~\cite{auto_nba}, resulting in 
\accuracyIncreaseANBA higher CNN accuracy and \fpsIncreaseANBA higher throughput, while enabling
\edpReductionANBA lower energy-delay product and \areaReductionANBA lower chip area.

\section*{Acknowledgments}

We performed the simulations presented in this article on computational resources managed and supported by Princeton Research Computing at Princeton University.

\bibliographystyle{ACM-Reference-Format}
\bibliography{biblio}

\end{document}